\documentclass[utf8]{frontiersSCNS} 

\usepackage{url,hyperref,lineno,microtype,subcaption}
\usepackage[onehalfspacing]{setspace}

\def\keyFont{\fontsize{8}{11}\helveticabold }
\def\firstAuthorLast{Järvelä {et~al.}} 
\def\Authors{Emilia Järvelä\,$^{1,*}$, Marco Berton\,$^{2,3}$ and Luca Crepaldi\,$^{4}$}
\def\Address{$^{1}$ European Space Agency (ESA), European Space Astronomy Centre (ESAC), Camino Bajo del Castillo s/n, 28692 Villanueva de la Ca\~nada, Madrid, Spain \\
$^{2}$ Finnish Centre for Astronomy with ESO - University of Turku, Vesilinnantie 5, FIN-20500, Turku, Finland \\
$^{3}$ Aalto University Metsähovi Radio Observatory, Metsähovintie 114, FIN-02450, Kylmälä, Finland \\
$^{4}$ Dipartimento di Fisica e Astronomia ``G. Galilei'', Universit\`a di Padova, Vicolo dell'Osservatorio 3, 35122, Padova, Italy
}
\def\corrAuthor{Emilia Järvelä}

\def\corrEmail{ejarvela@sciops.esa.int}

\begin{document}
\onecolumn
\firstpage{1}

\title[Radio spectral index maps of NLS1s]{Narrow-line Seyfert 1 galaxies with absorbed jets -- insights from radio spectral index maps}

\author[\firstAuthorLast ]{\Authors} 
\address{} 
\correspondance{} 

\extraAuth{}

\maketitle

\begin{abstract}

\section{}

Narrow-line Seyfert 1 (NLS1) galaxies are active galactic nuclei (AGN) believed to be in the early stages of their evolution. A fraction of them have been found to host relativistic jets. Due to the lack of large-scale diffuse radio emission they are believed to be experiencing one of their first activity cycles, and can offer us an opportunity to study the early evolution of more powerful AGN, such as radio galaxies and flat-spectrum radio quasars. Recently, a group of intriguing jetted NLS1s was discovered: based on high radio frequency data they host relativistic jets, but in the JVLA observations they all showed steep radio spectra at least up to 9.0~GHz, indicating very strong absorption at these frequencies. In this paper we study a subset of these sources in detail by employing spatially resolved radio spectral index maps at central frequencies of 1.6, 5.2, and 9.0~GHz. With spectral index maps we can disentangle the different radio emission components over the radio-emitting region, and get insights into the production mechanisms of radio emission. In addition, we study their host galaxies in relation to the radio emission to investigate if the host can provide us additional information regarding the origin of the radio emission, or the launching mechanism of the jets. It is fascinating how different the sources studied are, and certainly more, especially wide frequency-range, and high-resolution observations will be needed to understand their history and current properties, such as the reason behind the extraordinary radio spectra.

\tiny
 \keyFont{ \section{Keywords:} narrow-line Seyfert 1 galaxies, active galactic nuclei, radio emission, host galaxies, absorbed jets} 
\end{abstract}

\newcommand{\kms}{km s$^{-1}$}
\newcommand{\apjs}{ApJS}
\newcommand{\apj}{ApJ}
\newcommand{\aap}{A\&A}
\newcommand{\aj}{AJ}
\newcommand{\mnras}{MNRAS}
\newcommand{\nat}{Nature}
\newcommand{\apjl}{ApJL}
\newcommand{\aapr}{A\&ARv}
\newcommand{\aaps}{A\&AS}

\section{Introduction}
\defcitealias{2020berton2}{B20}

Narrow-line Seyfert 1 (NLS1) galaxies are a class of young active galactic nuclei (AGN) that are extremely interesting from the point of view of AGN evolution \citep{2000mathur1}. In their optical spectrum the permitted lines are of comparable width to their forbidden lines, and they are defined based on this property: by definition, the full-width at half maximum of the broad H$\beta$, FWHM(H$\beta$) $<$ 2000~\kms \citep{1985osterbrock1}. In addition the flux ratio [O~III] / H$\beta <$ 3 is required to ensure that we have a direct view of the broad-line region \citep[BLR,][]{1985osterbrock1}. NLS1s also often show strong Fe~II multiplets \citep{1989goodrich1}, but this feature is not part of the official classification criteria \citep{2016cracco1}.

The narrow FWHM of broad H$\beta$ is usually interpreted to be caused by low rotational velocity around a low- or intermediate mass supermassive black hole (generally $<10^8$ M$_\odot$, \citealp{2011peterson1}), in comparison to, for example, broad-line Seyfert 1 (BLS1) galaxies. Since the bolometric luminosities of NLS1s are comparable to those of BLS1s it implies that NLS1s are accreting close to or even above the Eddington limit \citep{1992boroson1}. An alternative explanation is that the narrow FWHM(H$\beta$) is caused by a flattened BLR seen face-on. In this case the permitted lines would appear narrow due to the lack of Doppler broadening, and the foreshortening of the velocity vector. In this scenario, the real central black holes masses can be considerably higher than what is derived from the BLR emission line properties, and NLS1s would be similar to BLS1s and other broad-line AGN \citep{2008decarli1}. However, inclination-independent observational properties of NLS1s, such as their host galaxy morphologies (\citealp[e.g.,][but see]{2001krongold1, 2006deo1, 2008anton1, 2016kotilainen1, 2018jarvela1, 2019berton1, 2020olguiniglesias1, 2021hamilton1} \citealp{2017dammando1, 2018dammando1}), and their different large-scale environments compared to BLS1s \citep{2017jarvela1} indicate that the black hole masses should be genuinely low. Detailed studies of some individuals, using, for example, reverberation mapping, support this hypothesis \citep{2014du1, 2016wang1, 2018du1, 2021berton1}, but more such studies are required to draw any definite conclusions. Assuming that the black hole masses in NLS1s are low, they may be at an early stage of AGN life cycle, and will eventually grow into fully developed broad-line AGN \citep{2000mathur1, 2000sulentic1, 2017fraixburnet1}. 

In AGN a host of different effluxes, launched by the central engine, are seen. These include collimated powerful relativistic jet, lower power non-relativistic jets, and non-collimated outflows and winds. A wide variety of AGN can launch low-power jets and outflows, but traditionally the most powerful relativistic jets were exclusively associated with the most massive supermassive black holes that reside in huge elliptical galaxies \citep{2000laor1}. This idea is turning out to be outdated as more observational evidence is gathered. For example, some NLS1s exhibit blazar-like properties, such as flat radio spectra\footnote{Flat spectrum has $\alpha >$ -0.5, when $S_{\nu} \propto \nu ^{\alpha}$ at frequency $\nu$.}, dominant radio emission, high brightness temperature, and prominent variability \citep{2006komossa1, 2008yuan1}. Indeed, the detection of the first NLS1 at gamma-rays, proving that it hosts relativistic jets, did not come as a surprise \citep{2009abdo2}. So far $\sim$20 NLS1s have been detected in gamma-rays, showing that, just like blazars and radio galaxies, NLS1s can harbour powerful relativistic jets \citep{2018romano1, 2020jarvela1, 2021rakshit1}. These NLS1s might be the progenitors of more powerful jetted AGN, like flat-spectrum radio quasars \citep{2015foschini1, 2017foschini1, 2020foschini1}.

However, only a fraction of NLS1s are blazar-like. In fact, most of them are very faint or not even detected in the radio band. It was thought that the radio emission in these sources would be mostly produces by star formation, or possibly by a less powerful form of AGN activity, such as low-power jets or outflows. Thus the discovery of strongly variable, almost Jy-level, emission at 37~GHz from some of these sources came as a surprise \citep{2018lahteenmaki1}. The properties of this emission can be explained only by the presence of relativistic jets. These sources became even more intriguing when they were observed with the Karl G. Jansky Very Large Array (JVLA) at 1.6, 5.2, and 9.0~GHz, and were found to show steep spectral indices at least up to 9.0~GHz, with no signs of jets, or even AGN activity \citepalias{2020berton2}. All the sources showed mJy- or sub-mJy-level flux densities, that could be explained by star formation alone, and also their spectral indices were in agreement with this scenario. To explaing the Jy-level detections at 37~GHz it seems evident that the spectrum needs to turn inverted above 9.0~GHz, but based on the JVLA observations it remained unclear whether these were young AGN with a semi-stationary peak at very high frequencies, like peaked-spectrum sources, or AGN with extreme variability. A connection between NLS1s and peaked-spectrum sources has already been made before \citep{2001oshlack1, 2006gallo1, 2006komossa1, 2014caccianiga1, 2016berton1, 2017berton1, 2017foschini1, 2017caccianiga1, 2021yao1, 2021odea1}. In this scenario the inverted spectrum would be caused by either synchrotron self-absorption or free-free absorption.

In this paper we aim to study the peculiar sources of \citet[][from now on B20]{2020berton2} in more detail. Our goal is to better characterise the radio emission at 1.6, 5.2, and 9.0~GHz to understand its origin, and investigate if any signs of the jets are detectable. We achieve this by producing and analysing the spatially resolved spectral index maps of these sources. Spectral index maps are superior to traditional spectral indices since they provide a spatially resolved view over the whole radio-emitting region, and enable us to identify different radio components, such as, optically thick and thin regions, and the radio emission production mechanisms. In addition to the radio data, we use any data and studies available in the literature to draw a more complete picture of these sources. We especially focus on their host galaxies in connection to the radio emission, as it can also help us understand the origin of the radio emission, and possibly the launching mechanism of the jets in these sources, as jetted NLS1s have found to often reside in interacting galaxies \citep{2008anton1,2018jarvela1}. The paper is organised as follows: in Sect.~\ref{sec:sample} we briefly overview the sample, in Sect.~\ref{sec:datared} we summarise the data reduction and the production of the radio and spectral index maps, in Sect.~\ref{sec:results} we present the results for individual sources, in Sect.~\ref{sec:discussion} we discuss our results and their implications, and finally, in Sect.\ref{sec:summary} we summarise this study. Throughout the paper, we use the standard $\Lambda$CDM cosmology, with $H_0$ = 70 km s$^{-1}$ Mpc$^{-1}$, and $\Omega_{\Lambda}$ = 0.73 \citep{2011komatsu1}. For spectral indices we adopt the convention of $S_{\nu} \propto \nu ^{\alpha}$ at frequency $\nu$.

\section{Sample}
\label{sec:sample}

\begin{figure}[h!]
\begin{center}
\includegraphics[width=18cm]{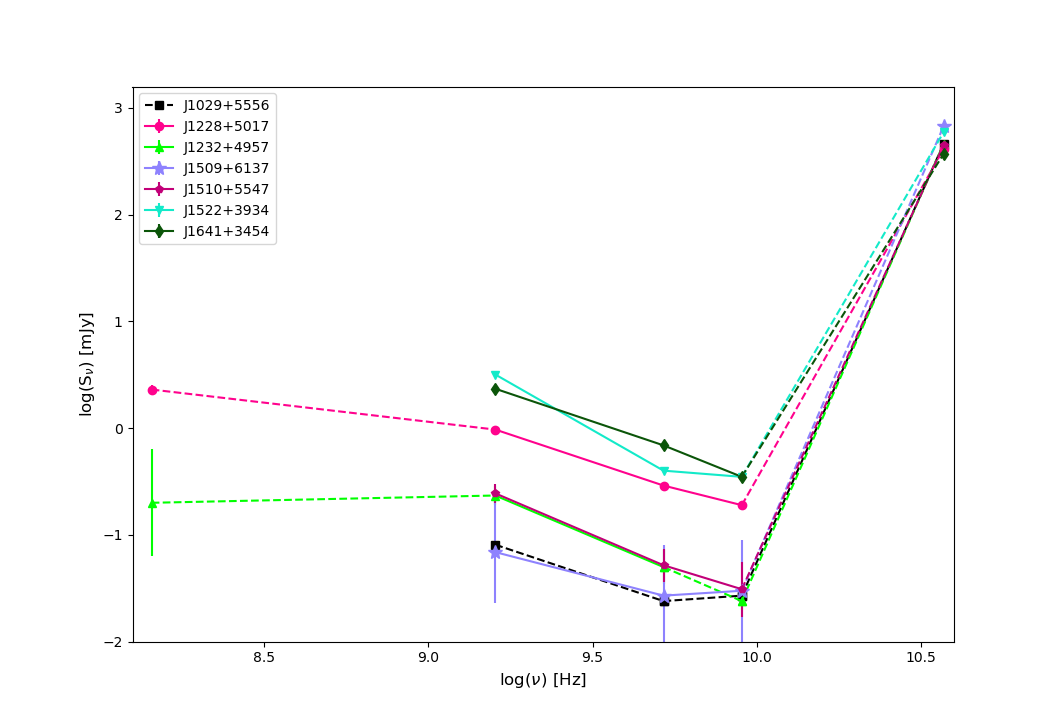}
\end{center}
\caption{Radio spectra of our sources from 144~MHz to 37~GHz. The JVLA data for J1228+5017, J1522+3934, and J1641+3454 from this paper, for other sources from \citetalias{2020berton2}. Simultaneous detections are connected with a solid line, archival detections and upper limits with a dashed line. The colours and symbols explained in the legend.}\label{fig:lcx-spectra}
\end{figure}

The sample includes seven NLS1s with extraordinary properties. These sources were either not detected at all or detected at very low flux densities in previous radio surveys, including Faint Images of the Radio Sky at Twenty-Centimeters (FIRST) \citep{1995becker1} and National Radio Astronomy Observatory (NRAO) Very Large Array (VLA) Sky Survey (NVSS) \citep{1998condon1}. However, they are included in the Metsähovi Radio Observatory NLS1 monitoring programme \citep{2017lahteenmaki1}, and were detected at 37~GHz at flux densities that strongly suggest that these sources host relativistic jets \citep{2018lahteenmaki1}. Furthermore, the radio emission proved to be highly variable, conclusively excluding any alternative mechanisms, for example, star formation processes, as the origin of the radio emission. These sources were observed with the JVLA in A-configuration in L-, C-, and X-bands, or, at 1.6, 5.2 and 9.0~GHz, respectively \citepalias{2020berton2}. Interestingly, all of them show steep radio spectra between 1.6 and 9.0~GHz as shown in Fig.~\ref{fig:lcx-spectra}. To explain the high flux densities seen at 37~GHz extreme flux density variability, a form of absorption, or a combination of them is needed. Since the required increase in the flux density would be two to three orders of magnitude, absorption seems a more plausible explanation. It is still unclear whether the main absorption mechanism is synchrotron self-absorption or free-free absorption. The JVLA data of these sources were studied in \citet{2020berton1}, but a more detailed analysis of the radio data, complemented by data at other wavelengths, might help us shed light on their nature.

\section{Data reduction and analysis}
\label{sec:datared}

The observations and calibration are described in detail in \citet{2020berton2}. However, for the analyses in this paper we used CASA version 5.6.2-3 due to its enhanced capabilities, for example, in producing spectral index maps. 

\subsection{Spectral index maps}
\label{sec:spindmaps}

The spectral index, or $\alpha$ maps were produced using the procedure described in Järvelä et al. (2021, submitted). We summarise the process in the next Sections.

\subsubsection{Cleaning}

We cleaned the sources using the multi-term multi-frequency synthesis, \texttt{mtmfs}, algorithm in CASA. It allows simultaneous fitting of a spectral index over the whole band-width and as a function of the position within the image using a simple power-law. The algorith performs the fit by modelling the spectrum of each flux component (pixel) by a Taylor series expansion about the reference frequency, $\nu_0$. In theory the number of Taylor terms is not limited, but in practise the quality of the data limits it. The number of resulting maps is equal to the number of Taylor terms. The first map, TT0, corresponds to the specific intensities at $\nu_0$ and is equal to the normal radio map. The second map, TT1, is defined so that $\alpha$ = TT1 / TT0. The third term, TT2, describes the spectral curvature and is defined so that $ \rm \beta = TT2 / TT0 -\alpha(\alpha-1)/2$. We chose to fit our sources with two Taylor terms to maximise the quality of the maps since the sources are faint, and the signal-to-noise (S/N) of the data is not very high. 

We produced the $\alpha$ maps in all bands to study possible changes in the spectral index within the frequency range of our observations. We were able to produce the maps for three sources, as the remaining four sources were either non-detections or did not have high enough S/N to produce reliable maps. The central frequencies of the $\alpha$ maps are 1.6, 5.2, and 9.0~GHz, and their band-widths are 1, 2, and 2~GHz, respectively. In addition to the $\alpha$ map, the \texttt{mtmfs} also produces a $\Delta \alpha$ error map that describes the empirical error estimate based on the errors of the TT0 and TT1 residual images. These maps require some post-processing steps, described below.

\subsubsection{Wide-band primary beam correction and 5$\sigma$ cut-off}
\label{sec:wbpbcorr}

The size of the primary beam depends on frequency and imposes its own spectral index onto the Taylor-coefficient images and the $\alpha$ map. We corrected this with the CASA task \texttt{widebandpbcor}, which computes a set of primary beams at given frequencies, calculates the Taylor-coefficient images representing the primary beam spectrum, performs the primary beam correction of the Taylor-coefficient images, and finally computes the primary beam corrected maps using the corrected Taylor-coefficient images.

This cannot take into account small variations during a single observations, such as the slightly changing shapes of the telescopes, and the primary beam rotation on the sky while tracking a source. However, we assume these errors to be insignificant. In addition, the primary beam errors increase with distance from the pointing centre, but according to \citet{2013bhatnagar1} the effects are minimal up to the half-power beam width (HPBW). Our sources are at all frequencies considerably smaller than the corresponding HPBW so this should not cause additional errors in the maps.

Furthermore, we masked pixels with a S/N $ < 5\sigma$, where $\sigma$ is the rms of the corresponding TT0 image of the source. In general, close to the edges of the emitting region the S/N is low, and the maps consistenly show extreme or clearly erroneous values. With this threshold we can remove unwanted noise from the $\alpha$ and $\Delta \alpha$ maps.

\subsubsection{$\Delta \alpha$ cut-off}
\label{sec:deltaalphacutoff}

After the primary beam correction the $\alpha$ maps can still show considerable variations, especially near the edges. Correspondingly, the $\Delta \alpha$ map shows very high errors in these regions, suggesting that the data quality might not have been good enough to accurately fit the spectral index. We cut off all the data that has $\Delta\alpha >$ 1. Even after this some erroneous values may remain, but we did not want to lose any information due to too high a threshold.

\subsubsection{Smoothing of the $\alpha$ and $\Delta\alpha$ maps}
\label{sec:alphasmoothing}

The Taylor coefficient maps are convolved with the clean beam, but as the $\alpha$ and $\Delta \alpha$ maps are derived from them using mathematical operations their final resolution is not the same as that of the Taylor coefficient maps. We thus smoothed the $\alpha$ and $\Delta \alpha$ maps using the parameters of the clean beam of each source at each frequency. This considerably decreases the small-scale variations of the $\alpha$ maps and the errors of the $\Delta \alpha$ maps, and produces $\alpha$ and $\Delta \alpha$ maps with a resolution similar to the normal radio maps of each source.

\subsubsection{Measurements}

We list the rms of the maps and the peak and the integrated flux densities in Table~\ref{tab:flux}. The rms was measured from an empty sky region off-source. We obtained the peak flux density by fitting a 2D Gaussian to the data, and the integrated flux density by summing up all the emission within the 3$\sigma$ contour. The peak flux density error is given by CASA when fitting a 2D Gaussian to the data, and the errors of the integrated flux densities were estimated with rms per beam $\times$ the square root of the area of emission expressed in beams. In addition, for each source we calculated an average spectral index over the whole region that the $\alpha$ map covers weighted with the surface brightness, and in the core in a region with a radius of 2~px. The total and core spectral indices are listed in Table~\ref{tab:spinds}.


\subsection{Traditional spectral indices}
\label{sec:tradspind}

To confirm the results of the $\alpha$ maps we also derived the traditional spectral indices of the three sources with existing $\alpha$ maps. We achieved this by dividing the observed band into two (spectral windows 0 to 7, and 8 to 15), cleaning these data sets separately, and smoothing the radio map with the higher resolution with the clean beam of the lower resolution map. We did this for each band, resulting in three in-band spectral indices for each source. We calculated the errors of the traditional spectral indices using standard error propagation, which results in rather high errors when using logarithmic values, and probably somewhat overestimates the final error. Since we are especially interested in whether any signs of a jet can be seen, we estimated the spectral indices using the peak flux densities, and they thus reflect the spectral index of the core. The results are shown in Table~\ref{tab:tradspind}.

\section{Results}
\label{sec:results}

Individual sources are discussed below. We focus on sources for which we were able to produce the $\alpha$ maps. In addition to the radio data, we gathered any information and data available in the literature. We also obtained the $i$-band host galaxy images and the colour images of the sources from the Panoramic Survey Telescope and Rapid Response System (Pan-STARRS). 

\subsection{J1228+5017}
\label{sec:j1228}

J1228+5017 is an NLS1 at $z$ = 0.262. It did not have any prior radio detections before the detections at Metsähovi at 37~GHz. Its integrated JVLA flux densities are very modest: (0.97 $\pm$ 0.04)~mJy, (0.29 $\pm$ 0.01)~mJy, and (0.19 $\pm$ 0.01)~mJy, at 1.6, 5.2, and 9.0~GHz, respectively, indicating a steep spectrum up to 9.0~GHz (Fig.~\ref{fig:lcx-spectra}). 

\begin{figure}[h!]
\begin{center}
\includegraphics[width=1\textwidth]{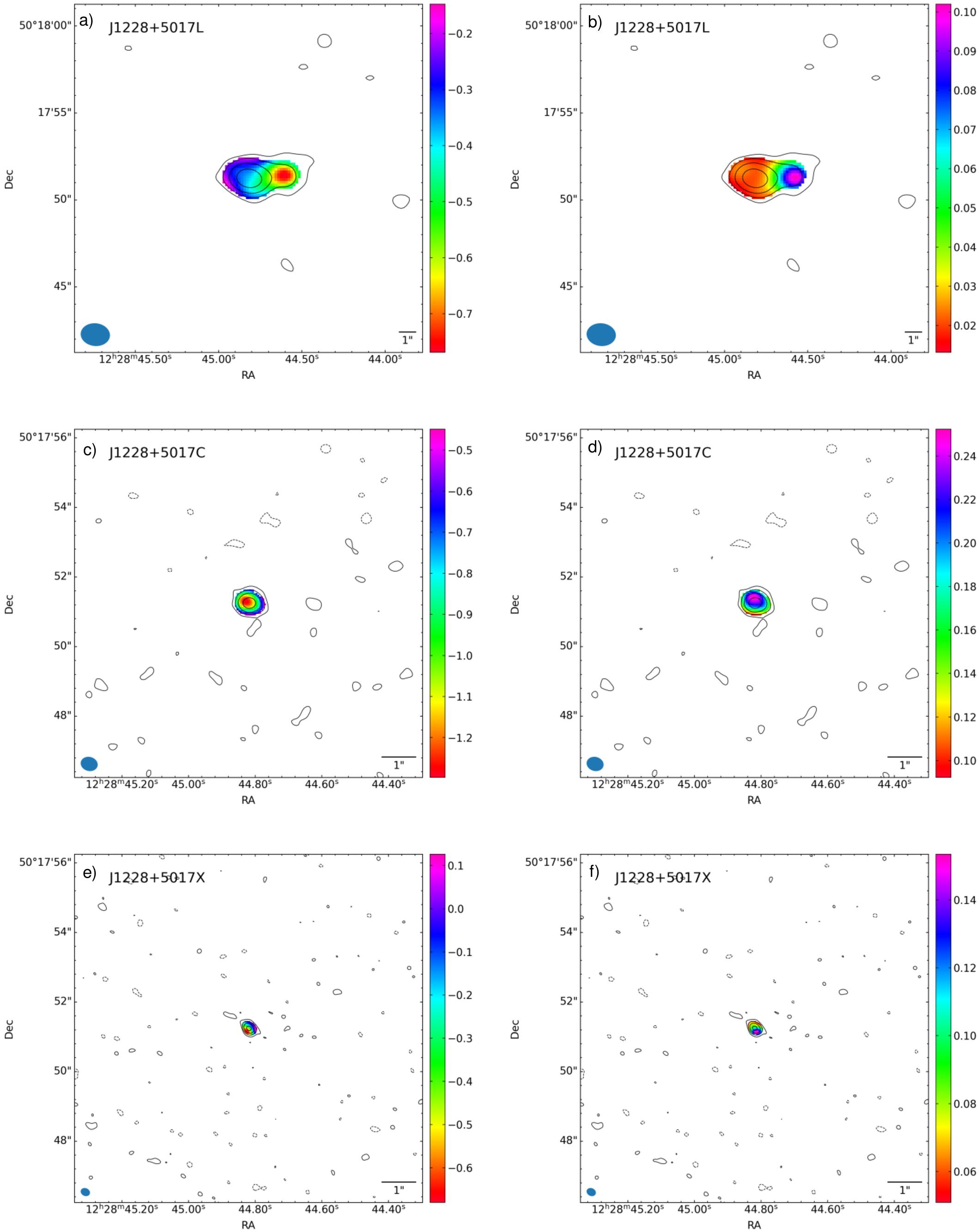}
\end{center}
\caption{\textit{a)} 1.6~GHz $\alpha$ map, rms = 20$\mu$Jy beam$^{-1}$, beam size 9.56 $\times$ 7.41~kpc; \textit{b)} 1.6~GHz $\Delta \alpha$ map; \textit{c)} 5.2~GHz $\alpha$ map, rms = 7.5$\mu$Jy beam$^{-1}$, beam size 2.83 $\times$ 2.31~kpc; \textit{d)} 5.2~GHz $\Delta \alpha$ map; \textit{e)} 9.0~GHz $\alpha$ map, rms = 6.5$\mu$Jy beam$^{-1}$, beam size 1.58 $\times$ 1.19~kpc \textit{f)} 9.0~GHz $\Delta \alpha$ map. Contour levels at -3, 3 $\times$ 2$^n$, $n \in$ [0, 3] in all figures. The scale is 5.657~kpc/arcsec.}\label{fig:J1228-radio}
\end{figure}

Its 1.6~GHz radio map (Fig.~\ref{fig:J1228-radio}a) shows an elongated structure toward west, with a maximum extent of 18.1~kpc. The 5.2 and 9.0~GHz maps(Figs.~\ref{fig:J1228-radio}c and e) do not show signs of this structure, suggesting a steep spectrum. The 1.6~GHz $\alpha$ map (Fig.~\ref{fig:J1228-radio}a) confirms this since the extended emission shows steep spectral indices, with values around -0.5 to -0.7, consistent with optically thin radio emission. Interestingly, the core seems quite flat, and indeed, its weighted spectral index is -0.38 $\pm$ 0.01. This result is supported by the traditional spectral index of -0.19 $\pm$ 0.19 at 1.6~GHz. At 5.2~GHz both methods give a steep spectral index, around -1.0. At 9.0~GHz the spectral index seems to flatten again, the core spectral index is -0.43 $\pm$ 0.03, and the traditional spectral index is -0.46 $\pm$ 0.91. However, taking into account the faintness of the source, we cannot be certain how reliable especially the 9.0~GHz spectral indices are, since a faint source might be barely detected in the highest frequency spectral windows, which can affect the fit, and a point-like source might suffer from the edge effects. Both of these are true for J1228+5017, so some caution should be exercised when intepreting the results. J1228+5017 was also detected in the Low Frequency Array (LOFAR) Two-metre Sky Survey (LoTSS) at 144~MHz with an integrated flux density of (2.3 $\pm$ 0.1)~mJy. This gives a 144~MHz to 1.6~GHz spectral index of -0.36 $\pm$ 0.04, consistent with the flat index seen at 1.6~GHz. It seems like the radio spectrum of J1228+5017 is flat at low frequencies, steepens above 1.6~GHz, might flatten again when reaching 9.0~GHz, and, to take into account the Metsähovi detections, probably turns inverted at some point above 9.0~GHz. Radio spectra with similar properties has been seen in some peaked-spectrum sources \citep[e.g., ][]{1990baum1, 2010hancock1, 2017callingham1}. This feature is explained as restarted activity manifesting as a peaked spectrum at higher frequencies, superimposed on a spectrum resulting from a period (or periods) of earlier activity and peaking at lower frequencies. Intermittent activity could be a plausible explanation since according to theoretical models young sources with high accretion rates are more prone to this kind of behaviour \citep{2009czerny1}. However, more observations over a wider frequency range, and also monitoring observations, are needed to confirm the shape of the spectrum, and the temporal evolution of this source.

The host galaxy of J1228+5017 was found to be a late-type galaxy (Björklund et al. in prep.). Also the host shows an elongated structure toward west and the 1.6~GHz radio emission traces this structure (Fig.~\ref{fig:hosts}a). This might be pure chance, or indicate that the extended radio emission actually originates from the host galaxy, in which case it would be produced by star formation related processes. However, the Pan-STARRS $i$-$r$-$g$ colour image (Fig.~\ref{fig:hosts}b) shows that the extended part is clearly red, unlike what would be expected in presence of strong star formation. The colour difference is so drastic that it is actually unclear if the blue and the red part belong to the same galaxy, if they are a pair of merging galaxies, or if the red source is a foreground or a background galaxy. The redshift of the red region needs to be confirmed with optical spectroscopy. It would also be interesting to measure the redshift of the galaxy on the west side of J1228+5017 as it might be a companion galaxy. Also the origin and the morphology of the radio emission needs to be studied in detail to understand what actually is the origin of the extended radio emission.

\subsection{J1522+3934}
\label{sec:j1522}

J1522+3934 is a nearby NLS1 at $z$ = 0.077. It was detected in FIRST with a flux density of 2.52~mJy, which is comparable to its optical flux density, implying that no enhanced activity was expected from the AGN. However, so far its maximum flux density observed at 37~GHz has been 1.43~Jy, unquestionably confirming the presence of relativistic jets \citep{2018lahteenmaki1}. Its integrated JVLA flux densities are (3.18 $\pm$ 0.07)~mJy, (0.40 $\pm$ 0.02)~mJy, and (0.35 $\pm$ 0.02)~mJy at 1.6, 5.2, and 9.0~GHz, respectively, consistent with a steep spectral index (Fig.~\ref{fig:lcx-spectra}).

\begin{figure}[h!]
\begin{center}
\includegraphics[width=1\textwidth]{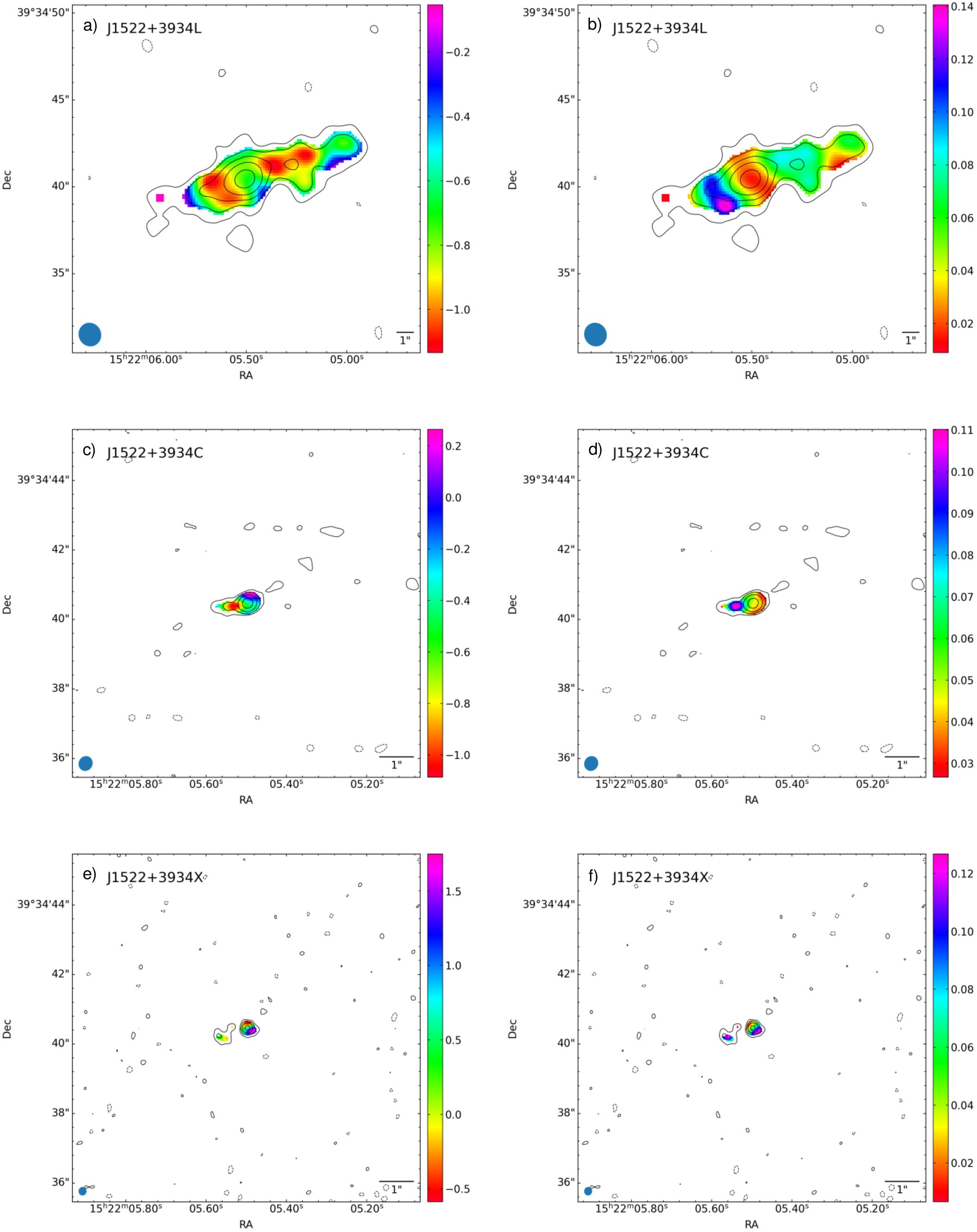}
\end{center}
\caption{\textit{a)} 1.6~GHz $\alpha$ map, rms = 17$\mu$Jy beam$^{-1}$, beam size 2.26 $\times$ 2.13~kpc; \textit{b)} 1.6~GHz $\Delta \alpha$ map; \textit{c)} 5.2~GHz $\alpha$ map, rms = 10$\mu$Jy beam$^{-1}$, beam size 0.73 $\times$ 0.66~kpc; \textit{d)} 5.2~GHz $\Delta \alpha$ map; \textit{e)} 9.0~GHz $\alpha$ map, rms = 7$\mu$Jy beam$^{-1}$, beam size 0.41 $\times$ 0.38~kpc \textit{f)} 9.0~GHz $\Delta \alpha$ map. Contour levels at -3, 3 $\times$ 2$^n$, $n \in$ [0, 4] in a) and b), and $n \in$ [0, 3] in c)-f). The scale is 1.650~kpc/arcsec.}\label{fig:J1522-radio}
\end{figure}

At 1.6~GHz it shows considerable symmetric extended emission in east/south-east - west/north-west direction (Fig.~\ref{fig:J1522-radio}a). The extended emission is barely detectable at 5.2 and 9.0~GHz (Figs.~\ref{fig:J1522-radio}c and e), indicating that its spectral index is quite steep. The total spectral index at 1.6~GHz is -0.80 and the core spectral index is -0.71. The $\alpha$ map (Fig.~\ref{fig:J1522-radio}a) shows regions of slightly steeper indices, but also the errors in these regions are higher (Fig.~\ref{fig:J1522-radio}b). Nevertheless, the core does not seem to be as steep as the extended emission regions. The traditionally estimated peak spectral index (-0.48 $\pm$ 0.30) is flatter than the core spectral index, but the results are in agreement within the error limits.

At 5.2~GHz the source shows only a slightly extended morphology toward east (Fig.~\ref{fig:J1522-radio}c). Interestingly, at this frequency, the source seems to be flat: the total spectral index is -0.42, and the core spectral index is -0.47. The traditional spectral index is in agreement with $\alpha$ = -0.55 $\pm$ 0.45. The 9.0~GHz map does not show considerable structures either, just a faint region of diffuse emission on the east side of the nucleus (Fig.~\ref{fig:J1522-radio}e). Also at this frequency the source seems flat: the total spectral index is 0.49, and the core spectral index is 0.35. The source is so faint at 9.0~GHz that these values are probably not totally reliable as edge and low S/N effects start to affect. However, the traditional spectral index also indicates a flat spectrum with $\alpha$ = -0.46 $\pm$ 1.08, but also in this case the result is affected by the poor S/N, resulting in huge errors. However, the 5.2 and 9.0~GHz behaviour together can be seen as a hint of the existence of the relativistic jets in the nucleus of the source.

The end-to-end extent of the emitting region at 1.6~GHz is 19.4~kpc, and when overlaid with its host galaxy $i$-band image from Pan-STARRS (Fig.~\ref{fig:hosts}c) it is evident that the radio emission is perpendicular to the disk of the host. The host was studied in detail in \citet{2018jarvela1} who found that it is a barred disk-like galaxy with a pseudo-bulge, and merging with another, possibly a dwarf, galaxy. The projected distance between the nuclei is 2.2~kpc and the radio emission is centred at the NLS1. The NLS1 nucleus seems to be strongly obscured in $i$-band compared to the near-infrared $Ks$-band (see Fig. 13 in \citealp{2018jarvela1}). Indeed, a prominent dust lane, probably obscuring the NLS1 nucleus, can be seen in the Pan-STARRS $g$ and $z$ -band colour image (Fig.~\ref{fig:hosts}d).

Taking into account the approximately 90 degree angle between the position angles of the host disk and the radio emission it seems improbable that star formation could be responsible for the radio emission. Also \citetalias{2020berton2}, using the strength of the mid-infrared emission as a proxy for the star formation, reached the conclusion that star formation alone is not enough to account for the 1.6~GHz radio emission. A more plausible explanation is that the AGN is responsible for the extended emission. Indeed, based on the 37~GHz observations we know that this source host relativistic jets, and taking into account the maximum flux density levels detected, exceeding 1~Jy, it would seem that beaming is needed, in which case we would be seeing the jet at a small angle. Also the flat spectral indices observed in the core support this. A small viewing angle is not consistent with the symmetrical extended morphology we see at 1.6~GHz, which implies a considerably larger inclination. One option is that the jets have changed direction, possibly by precession, or alternatively, they have turned off and restarted with a different viewing angle \citep{2017hernandezgarcia1}. There is no evidence of the S-shaped morphology usually seen in jet precession cases \citep{1978ekers1,1985parma1}. In fact, the morphology does not seem perturbed at all. Furthermore, the radio spectrum that remains steep at least up to 9.0~GHz, and seems to turn very inverted above that, supports the intermittency scenario as such extreme spectral indices are usually seen in kinematically young, absorbed jets, such as in peaked sources \citep[PS, e.g.,][]{2021odea1}. J1522+3934 is not a traditional peaked source since they usually exhibit steep core emission at kpc scales. Instead, J1522+3934 seems to be more core-dominated, adding further proof to the scenario in which it is seen at a small angle. \citet{2009czerny1} hypothesise that in young sources with high accretion rates, such as NLS1s, radiation pressure instabilities of the accretion disk can result in intermittent activity. They estimate that the activity periods last 10$^3$-10$^4$~yr and are separated by periods of 10$^4$-10$^6$~yr. In this scenario the extended emission would be a relic of the past activity period, and not replenished anymore if the jet direction has changed. However, it does not show extraordinarily steep spectral indices, as seen in the case of Mrk~783 \citep{2017congiu1,2020congiu1}, which might indicate that it is not very old. This is not unheard of since such sources have been observed before \citep[e.g., ][]{2002dennettthorpe1}.

Re-orientation of the jets, aligned with the spin axis of the black hole, can happen when some incident perturbs the dynamics of the very innermost region of an AGN, resulting in a change of the spin axis. Such an event can be, for example, a black hole - black hole merger, or an accretion period of matter with an angular momentum axis different to the current system axis, which can re-orient the accretion disk, and finally lead to a realignment of the black hole spin axis \citep{2002dennettthorpe1, 2012gopalkrishna1}. These events are often related to galaxy-galaxy mergers or interaction, which could be the culprit also in case of J1522+3934 since it resides in a merging system. More observations are certainly needed to investigate the properties of this intriguing source in more detail.




\subsection{J1641+3454}
\label{sec:j1641}

\begin{figure}[h!]
\begin{center}
\includegraphics[width=1\textwidth]{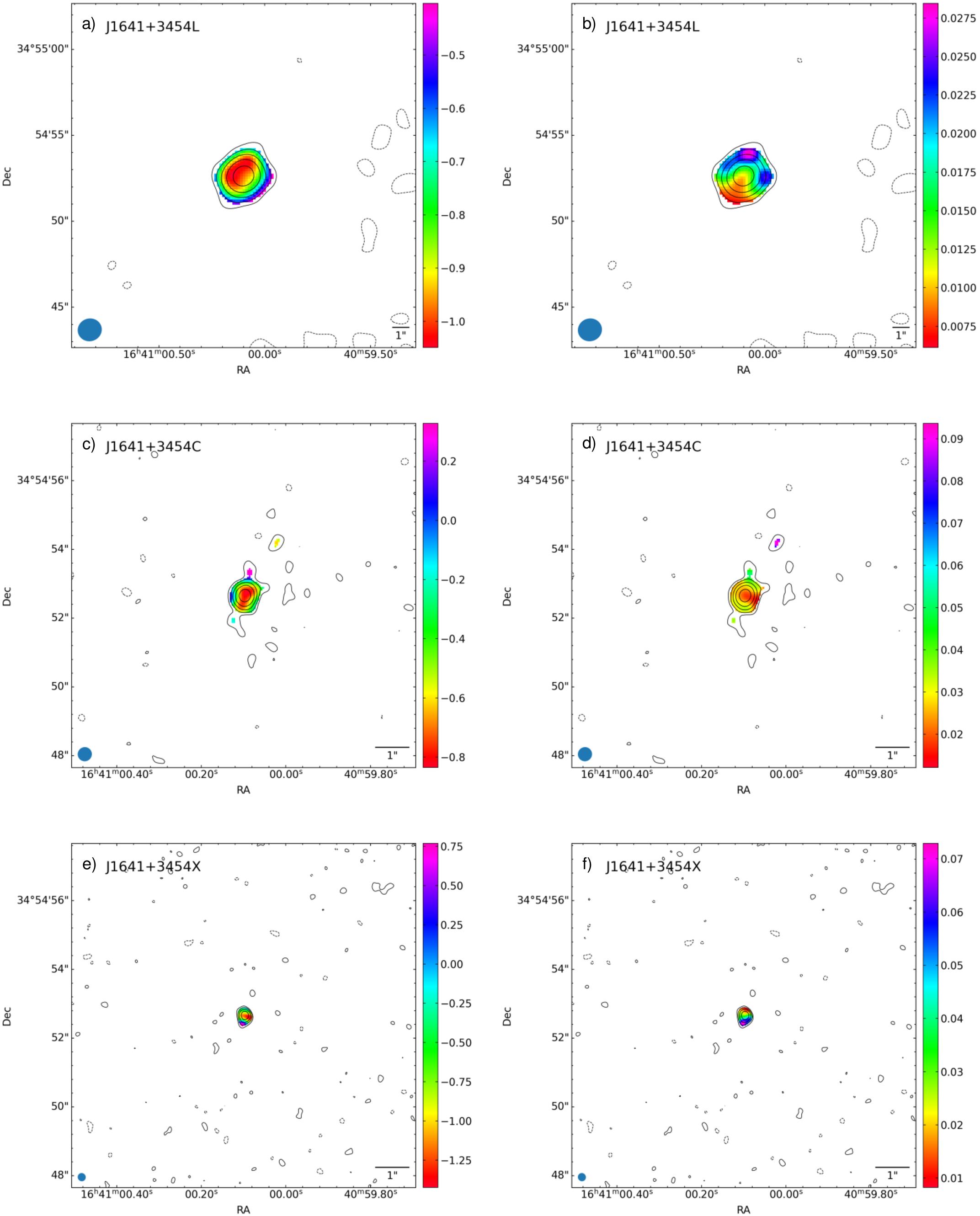}
\end{center}
\caption{\textit{a)} 1.6~GHz $\alpha$ map, rms = 25$\mu$Jy beam$^{-1}$, beam size 4.96 $\times$ 4.64~kpc; \textit{b)} 1.6~GHz $\Delta \alpha$ map; \textit{c)} 5.2~GHz $\alpha$ map, rms = 7$\mu$Jy beam$^{-1}$, beam size 1.48 $\times$ 1.44~kpc; \textit{d)} 5.2~GHz $\Delta \alpha$ map; \textit{e)} 9.0~GHz $\alpha$ map, rms = 7$\mu$Jy beam$^{-1}$, beam size 0.84 $\times$ 0.81~kpc \textit{f)} 9.0~GHz $\Delta \alpha$ map. Contour levels at -3, 3 $\times$ 2$^n$, $n \in$ [0, 4] in a)-d), and $n \in$ [0, 3] in e) and f). The scale is 3.518~kpc/arcsec.}\label{fig:J1641-radio}
\end{figure}

\begin{figure}[h!]
\begin{center}
\includegraphics[width=0.8\textwidth]{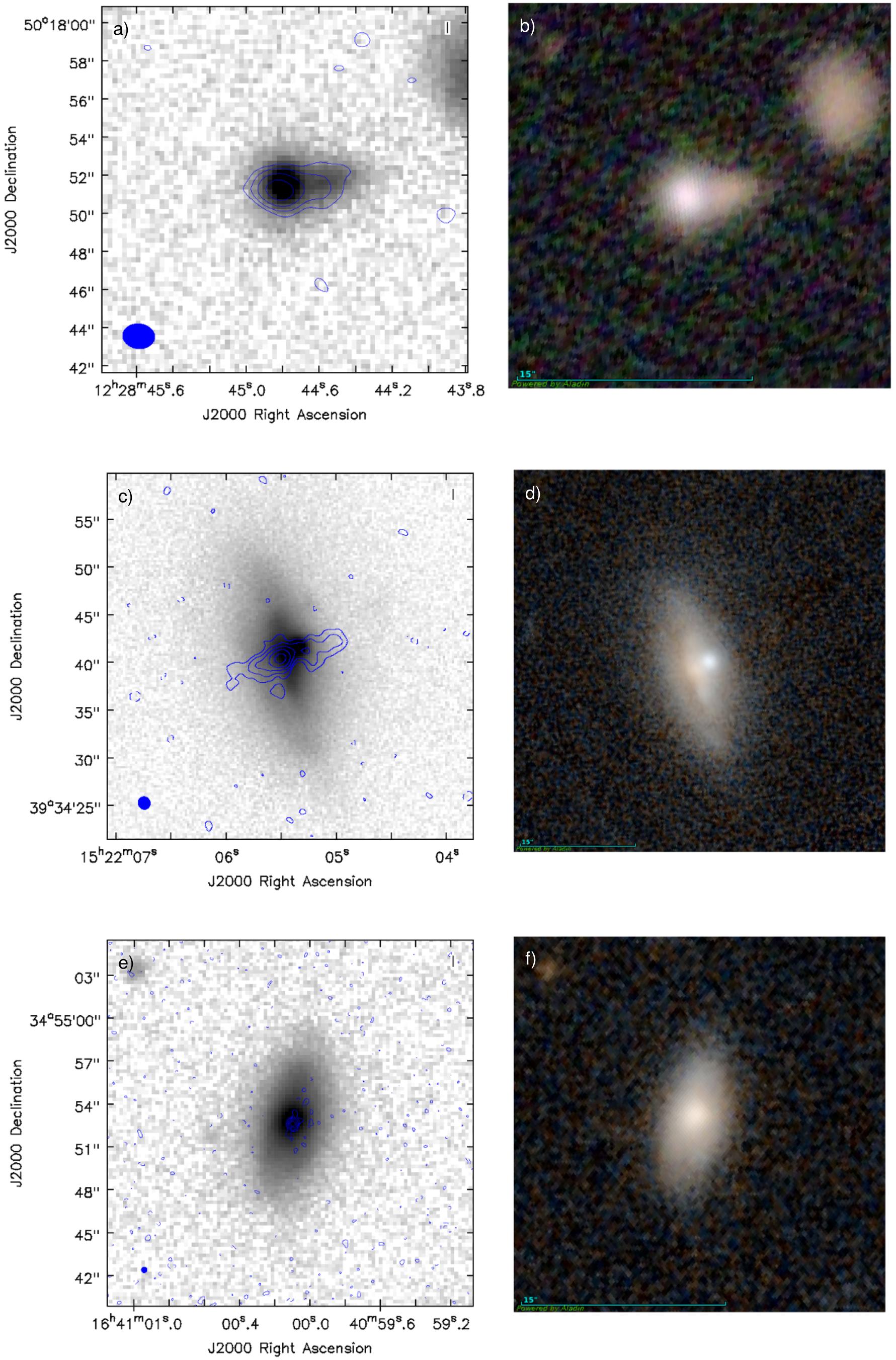}
\end{center}
\caption{\textit{a)} J1228+5017 1.6~GHz radio map overlaid with the $i$-band Pan-STARRS image; \textit{b)} J1228+5017 Pan-STARRS $i$-$r$-$g$ colour image; \textit{c)} J1522+3934 1.6~GHz radio map overlaid with the $i$-band Pan-STARRS image; \textit{d)} J1522+3934 Pan-STARRS $z$-$g$ colour image; \textit{e)} J1641+3454 5.2~GHz radio map overlaid with the $i$-band Pan-STARRS image; \textit{f)} J1641+3454 Pan-STARRS $z$-$g$ colour image.} \label{fig:hosts}
\end{figure}

J1641+3454 is an NLS1 at a redshift of 0.164. It was detected in FIRST with an integrated flux density of 2.69~mJy, so the radio emission slightly dominates over the optical emission. However, 37~GHz detections at Metsähovi, and later a gamma-ray detection by \textit{Fermi} \citep{2017lahteenmaki1} confirmed the presence of most likely relativistics jets in this source. The JVLA measurements do not reflect this at all since its integrated flux densities are (2.34 $\pm$ 0.06)~mJy, (0.69 $\pm$ 0.02)~mJy, and (0.35 $\pm$ 0.01)~mJy, at 1.6, 5.2, and 9.0~GHz, respectively (Fig.~\ref{fig:lcx-spectra}). Its 1.6~GHz radio map (Fig.~\ref{fig:J1641-radio}a) shows diffuse extended emission around the centre. At 5.2~GHz the emission resolves to north and southward extended parts, surrounded by diffuse-looking patches (Fig.~\ref{fig:J1641-radio}c). At 9.0~GHz there are no traces of the extended emission (Fig.~\ref{fig:J1641-radio}e). The $\alpha$ maps ((Fig.~\ref{fig:J1641-radio}a, c, e) as well as the traditional spectral indices agree that its radio emission has a steep spectral index at all observed frequencies, consistent with optically thin emission, or even steeper.

Estimating the star formation produced radio emission using mid-infrared measurements \citetalias{2020berton2} found that it is somewhat higher than what star formation should produce. There are no clear signs of the nuclear activity, but the source is slightly elongated in the north-south direction at 5.2 and 9.0~GHz maps, which could be due to the nuclear activity. On the other hand, the patchy emission around the core at 5.2~GHz resembles diffuse morphology often associated with star formation. In NLS1s high-level AGN activity, as well as nuclear star formation, can, and often do co-exist \citep[][Järvelä et al. 2021, submitted]{2010sani1, 2015caccianiga1}. The host (Figs.\ref{fig:J1641-radio}e and f) was found to be a barred late-type galaxy with a pseudo-bulge \citep{2020olguiniglesias1}, so the presence of star formation could be expected. It is interesting that unlike J1228+5017 and J1522+3923, J1641+3454 does not show any signs of flattening even at 9.0~GHz. More observations are required before we can determine whether it is similar to peaked-spectrum sources with a semi-stationary spectrum peaking at high frequency, a source with extreme variability, or possibly a combination of the two.

\section{Discussion}
\label{sec:discussion}

\subsection{Origin of the radio emission}

Based on \citetalias{2020berton2} the radio emission of these sources could be almost entirely explained with the star formation processes in their host galaxies. However, the spectral index maps revealed details, such as flat cores, not seen in the normal radio maps, and were able to give us a better view of the different mechanisms in play. Also the examination of the host galaxies in relation to the radio maps proved fruitful, especially in case of J1522+3934. Based on these new analyses it seems like J1641+3454 is the only source whose radio emission could be explained by star formation only. Both, J1228+5017 and J1522+3934 show a flat core, that can be consider as a certain sign of AGN activity in form of a jet or at least a jet base, at least at some frequency. It cannot be determined whether all the emission in J1228+5017 has an AGN origin, but it seems to be the predominant source. Interestingly, based on the 90 degree mismatch between the host and the extended radio emission in J1522+3934, and its flat core, it seems safe to assume that the AGN is the origin of almost all radio emission seen in this source. 

These results are somewhat surprising, taking into account the faintess of the emission, and their steep spectral indices measured in \citetalias{2020berton2}. This highlights the necessity to study AGN as individuals, as the general properties of these sources before the 37~GHz ad JVLA observations pointed to very different classification compared to what we have learned based on more detailed studies. These results also demostrate that spatially resolved spectral index maps can be utilised to examine the sources beyond what can be achieved with normal radio maps.

\subsection{Intermittent activity}

Two of the three sources show possible signs of intermittent activity. As discussed before in Sects.~\ref{sec:j1228} and \ref{sec:j1522}, this is not unexpected behaviour in young sources, accreting at high rates \citep{2009czerny1}. Examples of extreme realignments or precession, that can result in similar-looking features, are seen among many classes of AGN \citep{2002dennettthorpe1,2017hernandezgarcia1,2020congiu1}. Naturally, these sources show a host of different properties, depending on the duration of their earlier and current activity periods, and if and how the jet direction has changed. In radio band these sources can be identified by peculiar radio morphologies and very steep spectral indices of the emission of the past activity cycles, or by spectra showing unusual shapes. 

In case of J1522+3934 the telltale sign is the discrepancy between the large-scale radio moprhology and the fact that the currect radio emission seems to be beamed. Since no signs of precession are seen we assume the jets have been turned off during the re-orientation, or it has happened very rapidly. The extended emission does not show signs of aging since its spectral index does not deviate significantly from that of optically thin jet emission. This indicates that the re-orientation must have happened quite a short time ago. Also the peaked-spectrum source -like radio spectrum of J1522+3934 suggests that it is currently a kinematically young source. It can be speculated that the close galaxy merger J1522+3934 resides in might be the culprit behind the re-alignement, probably by means of an accretion episode of matter with a drastically different angular momentum vector than the earlier system axis. The case of J1228+5017 is not so clear, but the supposedly double-humped shape of its radio spectrum indicates that there might be some large-scale relic emission present at lower frequencies. Both these sources are extremely interesting, and, if confirmed, bring a valuable addition to the impressive group of AGN with intermittent activity.

\subsection{Host galaxies}

It has been found in earlier studies that NLS1s with relativistic jets seems to reside in interacting host galaxies more often than their non-jetted counterparts \citep[e.g., ][]{2008anton1, 2017jarvela1, 2003crenshaw1, 2007ohta1, 2011orbandexivry1}. However, we remark here that all the studied non-jetted NLS1s are at very low redshifts \citep[e.g.,][]{2003crenshaw1,2011orbandexivry1,2007ohta1} , whereas the jetted NLS1s tend to lie at higher redshifts \citep[e.g.,][]{2018romano1}, so the role of the cosmic evolution is unclear. Either way, observations of the hosts of non-jetted NLS1s at high redshifts are necessary to make any solid conclusions. It should be also noted that the claim of their non-jettedness is usually based on their low-frequency radio properties, and, for example, this paper demonstrates why that is not an advisable way to classify sources, let alone draw any conclusions about their properties.

All this kept in mind, our results do not lean either way. J1522+3934 is a late-type galaxy clearly in a merger, and it has also been confirmed with spectroscopy \citep{2017jarvela1}. Taking into account the small projected separation of the nuclei it seems to be an evolved merger, and the galaxies can be expected to affect, for example, the gas dynamics of each other. Like mentioned before, this can be the cause behind the jet re-oriention in J1522+3934. On the other hand, J1641+3454 was carefully studied in optical band in \citet{2020olguiniglesias1}, and they found it to be a pristine late-type galaxy, with no signs of interaction. Though minor mergers can go unnoticed when studying only the optical morphology, and require more sensivite methods, for example, spatially resolved spectroscopy to detect them \citep{2018longinotti1}. Nothing definite can be said about the status of J1228+5017 before further observations to determine if the sources optically close to it are at the same redshift. It is noteworthy that all these sources host powerful jets, but reside in late-type galaxies, directly contradicting the traditional jet paradigm \citep{2000laor1}. So much evidence has gathered against it after it was formulated that we should finally let go of it, and see the host galaxy as well as the black hole mass as one factor affecting the evolution and properties of AGN, but definitely not the determining one \citep{2020foschini1}.

\subsection{Absorption or extreme variability?}

There is a stark dissonance between the low- and high-frequency properties of these sources: at least up to 9.0~GHz they show generally steep radio spectra, and are very faint, at mJy or $\mu$Jy levels, on the other hand, at 37~GHz their average flux densities are of the order of $\sim$500~mJy, exceeding 1~Jy. To explain these properties some form of absorption, extreme variability, or both need to be present. It seems unlikely that only variability could explain this behaviour. For example, if J1522+3934 had a flat spectrum from 9.0 to 37~GHz, its flux density would have to 4000-fold to explain the peak flux densities seen at 37~GHz. The beam size difference between the JVLA and Metsähovi can account for some of the discrepancy but hardly this much. Furthermore, also the JVLA observations cover the whole host galaxy, so the source of the variable radio emission would need to be outside the host, or very extended, both of which seem very improbable.

It seems inevitable that the spectrum needs to turn inverted above 9.0~GHz, for which a form of absorption is needed. With our current data it remains unclear whether the spectral indices are extreme enough to require free-free absorption (FFA, with a theoretical maximum spectral index of 4), or if synchrotron self-absorption would be enough (SSA, with a maximum spectral index of $\sim$2.5 in a simplified case). Either way, these sources resemble peaked-spectrum sources, such as compact steep-spectrum and gigahertz-peaked sources, that are known to be AGN with kinematically young jets \citep{2021odea1}, that are absorbed either via FFA or SSA. However, some properties set them apart from the general peaked-spectrum source population: flat core spectral indices in kpc-scale observations, with the exception of J1641+3454, and the high-frequency variability. Generally peaked-spectrum sources show very moderate variability, and flat cores are detected in them only with very high resolution observations. Since our sources seem to be more core-dominated and show signs of beaming, like the 37~GHz radio flares, an alternative explanation might be needed. To avoid unphysical flux density increase an absorbed component is likely present above 9.0~GHz, but it cannot account for the high-amplitude, rapid flares, thus it seems plausible that the jet-base and possibly the innermost parts of the jet are not absorbed. Also the flat core spectra supports this scenario. Such a geometry can be achieved if we assume that the high-frequency peaked component is caused by FFA in a limited space, for example, in shocked ambient clouds in front of the jet head, and we see the jet at an optimal angle where beaming is still effective, but the angle is large enough so that the core/inner jet emission does not get absorbed by the FFA screen. Another mechanism that produces flat, $\alpha \sim$ -0.1, radio emission in galaxies is the free-free emission of ionised gas. In principle, if the jet is absorbed by FFA, we have ample ionised gas in the jet - interstellar medium shock front, that can produce free-free emission. However, according to studies this emission is swamped by the synchtroton emission in the shock, and not prominent \citep{1990contini1}. However, at this point our knowledge of these sources is so limited that we can only guess what their real nature is. In any case, they are extraordinary sources with distinctive properties, and should definitely be studied in detail in the future.

\section{Summary and conclusions}
\label{sec:summary}

In this paper we studied a sample of NLS1 galaxies exhibiting strongly variable emission at 37~GHz in single dish observations, but showing steep spectral indices up to 9.0~GHz in JVLA observations, indicating strong absorption and/or extreme variability. Our aim was to study their 1.6, 5.2, and 9.0~GHz radio emission in more detail, to investigate the origin of their radio emission, and check if any signs of the jet activity are detectable. Our main tool were the spatially resolved spectral index maps of each source at all observed frequencies. We were able to produce these maps for three sources, J1228+5017, J1522+3934, and J1641+3454. All of these sources proved to be quite different in their properties. The only common factor seems to be that they are hosted by late-type galaxies, albeit J1522+3934 is in a merger, and J1228+5017 might be interacting as well. J1228+5017 and J1522+3934 also show signs of the AGN activity in form of flat core spectral indices, and in J1522+3934 the AGN seems to be the predominant source of radio emission. Both these NLS1s show some indication of past activity periods and re-started activity, and, furthermore, the jet direction of J1522+3934 might have changed. J1641+3454 does not show any signs of the jets as its total and core spectral indices remain steep up to 9.0~GHz, and its radio emission could be explained with star formation processes. However, a past gamma-ray detection strongly implies that the jets are present, but they must be heavily absorbed below 9.0~GHz. 

NLS1s are complicated sources, where often an interplay between the AGN, the various phenomena related to it, and the host galaxy can be seen. As such they, however, offer us an unprecedented view of the early stages of the evolution of powerful AGN, like flat-spectrum radio quasars, without the need to go to high redshifts. They can offer us clues about the circumstances required to trigger the jets, as well as the long-term behaviour of jetted systems, including intermittent activity. For now our knowledge of NLS1s is still very limited, and even more so with the peculiar individuals investigated in this paper. Further observations, especially simultaneous observations of the whole radio spectrum, more observations at high radio frequencies, high-resolution radio imaging, and spatially resolved spectroscopy will be needed to understand the emission and absorption mechanisms in play in these sources.

\section*{Conflict of Interest Statement}
The authors declare that the research was conducted in the absence of any commercial or financial relationships that could be construed as a potential conflict of interest.

\section*{Author Contributions}
Conceptualisation, analysis and writing, E.J.; analysis and writing, M.B.; analysis and writing L.C. All authors have read and agreed to the published version of this manuscript.

\section*{Funding}
This research received no external funding.

\section*{Acknowledgments}

E.J. is a current ESA research fellow. The authors thank E.C. for his help in the data visualisation. This paper is based (in part) on results obtained with LOFAR equipment. LOFAR \citep{2013vanhaarlem1} is the Low Frequency Array designed and constructed by ASTRON. The National Radio Astronomy Observatory is a facility of the National Science Foundation operated under cooperative agreement by Associated Universities, Inc. The Pan-STARRS1 Surveys (PS1) and the PS1 public science archive have been made possible through contributions by the Institute for Astronomy, the University of Hawaii, the Pan-STARRS Project Office, the Max-Planck Society and its participating institutes, the Max Planck Institute for Astronomy, Heidelberg and the Max Planck Institute for Extraterrestrial Physics, Garching, The Johns Hopkins University, Durham University, the University of Edinburgh, the Queen's University Belfast, the Harvard-Smithsonian Center for Astrophysics, the Las Cumbres Observatory Global Telescope Network Incorporated, the National Central University of Taiwan, the Space Telescope Science Institute, the National Aeronautics and Space Administration under Grant No. NNX08AR22G issued through the Planetary Science Division of the NASA Science Mission Directorate, the National Science Foundation Grant No. AST-1238877, the University of Maryland, Eotvos Lorand University (ELTE), the Los Alamos National Laboratory, and the Gordon and Betty Moore Foundation. Funding for the Sloan Digital Sky Survey (SDSS) has been provided by the Alfred P. Sloan Foundation, the Participating Institutions, the National Aeronautics and Space Administration, the National Science Foundation, the U.S. Department of Energy, the Japanese Monbukagakusho, and the Max Planck Society. The SDSS Web site is http://www.sdss.org/. The SDSS is managed by the Astrophysical Research Consortium (ARC) for the Participating Institutions. The Participating Institutions are The University of Chicago, Fermilab, the Institute for Advanced Study, the Japan Participation Group, The Johns Hopkins University, the Korean Scientist Group, Los Alamos National Laboratory, the Max-Planck-Institute for Astronomy (MPIA), the Max-Planck-Institute for Astrophysics (MPA), New Mexico State University, University of Pittsburgh, University of Portsmouth, Princeton University, the United States Naval Observatory, and the University of Washington. This research has made use of the NASA/IPAC Extragalactic Database (NED), which is operated by the Jet Propulsion Laboratory, California Institute of Technology, under contract with the National Aeronautics and Space Administration. This research has made use of the SIMBAD database, operated at CDS, Strasbourg, France.

\section*{Data Availability Statement}
The datasets analysed for this study can be found in the NRAO Archive, \newline
https://archive.nrao.edu/archive/advquery.jsp.

\bibliographystyle{frontiersinSCNS_ENG_HUMS} 
\bibliography{artikkeli.bib}

\begin{thebibliography}{70}
\providecommand{\natexlab}[1]{#1}
\expandafter\ifx\csname urlstyle\endcsname\relax
  \providecommand{\doi}[1]{doi:\discretionary{}{}{}#1}\else
  \providecommand{\doi}{doi:\discretionary{}{}{}\begingroup
  \urlstyle{rm}\Url}\fi
\providecommand{\selectlanguage}[1]{\relax}
\providecommand{\bibAnnoteFile}[1]{%
  \IfFileExists{#1}{\begin{quotation}\noindent\textsc{Key:} #1\\
  \textsc{Annotation:}\ \input{#1}\end{quotation}}{}}
\providecommand{\bibAnnote}[2]{%
  \begin{quotation}\noindent\textsc{Key:} #1\\
  \textsc{Annotation:}\ #2\end{quotation}}

\bibitem[{{Abazajian} et~al.(2009){Abazajian}, {Adelman-McCarthy},
  {Ag{\"u}eros}, {Allam}, {Allende Prieto}, {An} et~al.}]{2009abazajian1}
{Abazajian}, K.~N., {Adelman-McCarthy}, J.~K., {Ag{\"u}eros}, M.~A., {Allam},
  S.~S., {Allende Prieto}, C., {An}, D., et~al. (2009).
\newblock {The Seventh Data Release of the Sloan Digital Sky Survey}.
\newblock \emph{\apjs} 182, 543-558.
\newblock \doi{10.1088/0067-0049/182/2/543}
\bibAnnoteFile{2009abazajian1}

\bibitem[{{Abdo} et~al.(2009){Abdo}, {Ackermann}, {Ajello}, {Axelsson},
  {Baldini}, {Ballet} et~al.}]{2009abdo2}
{Abdo}, A.~A., {Ackermann}, M., {Ajello}, M., {Axelsson}, M., {Baldini}, L.,
  {Ballet}, J., et~al. (2009).
\newblock {Fermi/Large Area Telescope Discovery of Gamma-Ray Emission from a
  Relativistic Jet in the Narrow-Line Quasar PMN J0948+0022}.
\newblock \emph{\apj} 699, 976--984.
\newblock \doi{10.1088/0004-637X/699/2/976}
\bibAnnoteFile{2009abdo2}

\bibitem[{{Ant{\'o}n} et~al.(2008){Ant{\'o}n}, {Browne}, and
  {March{\~a}}}]{2008anton1}
{Ant{\'o}n}, S., {Browne}, I.~W.~A., and {March{\~a}}, M.~J. (2008).
\newblock {The colour of the narrow line Sy1-blazar 0324+3410}.
\newblock \emph{\aap} 490, 583--587.
\newblock \doi{10.1051/0004-6361:20078926}
\bibAnnoteFile{2008anton1}

\bibitem[{{Baum} et~al.(1990){Baum}, {O'Dea}, {Murphy}, and {de
  Bruyn}}]{1990baum1}
{Baum}, S.~A., {O'Dea}, C.~P., {Murphy}, D.~W., and {de Bruyn}, A.~G. (1990).
\newblock {0108+388 : a compact double source with surprising properties.}
\newblock \emph{\aap} 232, 19
\bibAnnoteFile{1990baum1}

\bibitem[{{Becker} et~al.(1995){Becker}, {White}, and {Helfand}}]{1995becker1}
{Becker}, R.~H., {White}, R.~L., and {Helfand}, D.~J. (1995).
\newblock {The FIRST Survey: Faint Images of the Radio Sky at Twenty
  Centimeters}.
\newblock \emph{\apj} 450, 559.
\newblock \doi{10.1086/176166}
\bibAnnoteFile{1995becker1}

\bibitem[{{Berton} et~al.(2020{\natexlab{a}}){Berton}, {Bj{\"o}rklund},
  {L{\"a}hteenm{\"a}ki}, {Congiu}, {J{\"a}rvel{\"a}}, {Terreran}
  et~al.}]{2020berton1}
{Berton}, M., {Bj{\"o}rklund}, I., {L{\"a}hteenm{\"a}ki}, A., {Congiu}, E.,
  {J{\"a}rvel{\"a}}, E., {Terreran}, G., et~al. (2020{\natexlab{a}}).
\newblock {Line shapes in narrow-line Seyfert 1 galaxies: a tracer of physical
  properties?}
\newblock \emph{Contributions of the Astronomical Observatory Skalnate Pleso}
  50, 270--292.
\newblock \doi{10.31577/caosp.2020.50.1.270}
\bibAnnoteFile{2020berton1}

\bibitem[{{Berton} et~al.(2016){Berton}, {Caccianiga}, {Foschini}, {Peterson},
  {Mathur}, {Terreran} et~al.}]{2016berton1}
{Berton}, M., {Caccianiga}, A., {Foschini}, L., {Peterson}, B.~M., {Mathur},
  S., {Terreran}, G., et~al. (2016).
\newblock {Compact steep-spectrum sources as the parent population of
  flat-spectrum radio-loud narrow-line Seyfert 1 galaxies}.
\newblock \emph{\aap} 591, A98.
\newblock \doi{10.1051/0004-6361/201628171}
\bibAnnoteFile{2016berton1}

\bibitem[{{Berton} et~al.(2019){Berton}, {Congiu}, {Ciroi}, {Komossa},
  {Frezzato}, {Di Mille} et~al.}]{2019berton1}
{Berton}, M., {Congiu}, E., {Ciroi}, S., {Komossa}, S., {Frezzato}, M., {Di
  Mille}, F., et~al. (2019).
\newblock {The Interacting Late-type Host Galaxy of the Radio-loud Narrow-line
  Seyfert 1 IRAS 20181-2244}.
\newblock \emph{\aj} 157, 48.
\newblock \doi{10.3847/1538-3881/aaf5ca}
\bibAnnoteFile{2019berton1}

\bibitem[{{Berton} et~al.(2017){Berton}, {Foschini}, {Caccianiga}, {Ciroi},
  {Congiu}, {Cracco} et~al.}]{2017berton1}
{Berton}, M., {Foschini}, L., {Caccianiga}, A., {Ciroi}, S., {Congiu}, E.,
  {Cracco}, V., et~al. (2017).
\newblock {An orientation-based unification of young jetted active galactic
  nuclei: the case of 3C 286}.
\newblock \emph{Frontiers in Astronomy and Space Sciences} 4, 8.
\newblock \doi{10.3389/fspas.2017.00008}
\bibAnnoteFile{2017berton1}

\bibitem[{{Berton} et~al.(2020{\natexlab{b}}){Berton}, {J{\"a}rvel{\"a}},
  {Crepaldi}, {L{\"a}hteenm{\"a}ki}, {Tornikoski}, {Congiu}
  et~al.}]{2020berton2}
{Berton}, M., {J{\"a}rvel{\"a}}, E., {Crepaldi}, L., {L{\"a}hteenm{\"a}ki}, A.,
  {Tornikoski}, M., {Congiu}, E., et~al. (2020{\natexlab{b}}).
\newblock {Absorbed relativistic jets in radio-quiet narrow-line Seyfert 1
  galaxies}.
\newblock \emph{\aap} 636, A64.
\newblock \doi{10.1051/0004-6361/202037793}
\bibAnnoteFile{2020berton2}

\bibitem[{{Berton} et~al.(2021){Berton}, {Peluso}, {Marziani}, {Komossa},
  {Foschini}, {Ciroi} et~al.}]{2021berton1}
{Berton}, M., {Peluso}, G., {Marziani}, P., {Komossa}, S., {Foschini}, L.,
  {Ciroi}, S., et~al. (2021).
\newblock {Hunting for the nature of the enigmatic narrow-line Seyfert 1 galaxy
  PKS 2004-447}.
\newblock \emph{arXiv e-prints} , arXiv:2106.12536
\bibAnnoteFile{2021berton1}

\bibitem[{{Bhatnagar} et~al.(2013){Bhatnagar}, {Rau}, and
  {Golap}}]{2013bhatnagar1}
{Bhatnagar}, S., {Rau}, U., and {Golap}, K. (2013).
\newblock {Wide-field wide-band Interferometric Imaging: The WB A-Projection
  and Hybrid Algorithms}.
\newblock \emph{\apj} 770, 91.
\newblock \doi{10.1088/0004-637X/770/2/91}
\bibAnnoteFile{2013bhatnagar1}

\bibitem[{{Boroson} and {Green}(1992)}]{1992boroson1}
{Boroson}, T.~A. and {Green}, R.~F. (1992).
\newblock {The emission-line properties of low-redshift quasi-stellar objects}.
\newblock \emph{\apjs} 80, 109--135.
\newblock \doi{10.1086/191661}
\bibAnnoteFile{1992boroson1}

\bibitem[{{Caccianiga} et~al.(2014){Caccianiga}, {Ant{\'o}n}, {Ballo},
  {Dallacasa}, {Della Ceca}, {Fanali} et~al.}]{2014caccianiga1}
{Caccianiga}, A., {Ant{\'o}n}, S., {Ballo}, L., {Dallacasa}, D., {Della Ceca},
  R., {Fanali}, R., et~al. (2014).
\newblock {SDSS J143244.91+301435.3: a link between radio-loud narrow-line
  Seyfert 1 galaxies and compact steep-spectrum radio sources?}
\newblock \emph{\mnras} 441, 172--186.
\newblock \doi{10.1093/mnras/stu508}
\bibAnnoteFile{2014caccianiga1}

\bibitem[{{Caccianiga} et~al.(2015){Caccianiga}, {Ant{\'o}n}, {Ballo},
  {Foschini}, {Maccacaro}, {Della Ceca} et~al.}]{2015caccianiga1}
{Caccianiga}, A., {Ant{\'o}n}, S., {Ballo}, L., {Foschini}, L., {Maccacaro},
  T., {Della Ceca}, R., et~al. (2015).
\newblock {WISE colours and star formation in the host galaxies of radio-loud
  narrow-line Seyfert 1}.
\newblock \emph{\mnras} 451, 1795--1805.
\newblock \doi{10.1093/mnras/stv939}
\bibAnnoteFile{2015caccianiga1}

\bibitem[{{Caccianiga} et~al.(2017){Caccianiga}, {Dallacasa}, {Ant{\'o}n},
  {Ballo}, {Berton}, {Mack} et~al.}]{2017caccianiga1}
{Caccianiga}, A., {Dallacasa}, D., {Ant{\'o}n}, S., {Ballo}, L., {Berton}, M.,
  {Mack}, K.-H., et~al. (2017).
\newblock {SDSSJ143244.91+301435.3 at VLBI: a compact radio galaxy in a
  narrow-line Seyfert 1}.
\newblock \emph{\mnras} 464, 1474--1480.
\newblock \doi{10.1093/mnras/stw2471}
\bibAnnoteFile{2017caccianiga1}

\bibitem[{{Callingham} et~al.(2017){Callingham}, {Ekers}, {Gaensler}, {Line},
  {Hurley-Walker}, {Sadler} et~al.}]{2017callingham1}
{Callingham}, J.~R., {Ekers}, R.~D., {Gaensler}, B.~M., {Line}, J.~L.~B.,
  {Hurley-Walker}, N., {Sadler}, E.~M., et~al. (2017).
\newblock {Extragalactic Peaked-spectrum Radio Sources at Low Frequencies}.
\newblock \emph{\apj} 836, 174.
\newblock \doi{10.3847/1538-4357/836/2/174}
\bibAnnoteFile{2017callingham1}

\bibitem[{{Condon} et~al.(1998){Condon}, {Cotton}, {Greisen}, {Yin}, {Perley},
  {Taylor} et~al.}]{1998condon1}
{Condon}, J.~J., {Cotton}, W.~D., {Greisen}, E.~W., {Yin}, Q.~F., {Perley},
  R.~A., {Taylor}, G.~B., et~al. (1998).
\newblock {The NRAO VLA Sky Survey}.
\newblock \emph{\aj} 115, 1693--1716.
\newblock \doi{10.1086/300337}
\bibAnnoteFile{1998condon1}

\bibitem[{{Congiu} et~al.(2017){Congiu}, {Berton}, {Giroletti}, {Antonucci},
  {Caccianiga}, {Kharb} et~al.}]{2017congiu1}
{Congiu}, E., {Berton}, M., {Giroletti}, M., {Antonucci}, R., {Caccianiga}, A.,
  {Kharb}, P., et~al. (2017).
\newblock {Kiloparsec-scale emission in the narrow-line Seyfert 1 galaxy Mrk
  783}.
\newblock \emph{\aap} 603, A32.
\newblock \doi{10.1051/0004-6361/201730616}
\bibAnnoteFile{2017congiu1}

\bibitem[{{Congiu} et~al.(2020){Congiu}, {Kharb}, {Tarchi}, {Berton},
  {Caccianiga}, {Chen} et~al.}]{2020congiu1}
{Congiu}, E., {Kharb}, P., {Tarchi}, A., {Berton}, M., {Caccianiga}, A.,
  {Chen}, S., et~al. (2020).
\newblock {The radio structure of the narrow-line Seyfert 1 Mrk 783 with VLBA
  and e-MERLIN}.
\newblock \emph{\mnras} 499, 3149--3157.
\newblock \doi{10.1093/mnras/staa3024}
\bibAnnoteFile{2020congiu1}

\bibitem[{{Contini} and {Viegas-Aldrovandi}(1990)}]{1990contini1}
{Contini}, M. and {Viegas-Aldrovandi}, S.~M. (1990).
\newblock {Continuum Radiation from Active Galactic Nuclei}.
\newblock \emph{\apj} 350, 125.
\newblock \doi{10.1086/168367}
\bibAnnoteFile{1990contini1}

\bibitem[{{Cracco} et~al.(2016){Cracco}, {Ciroi}, {Berton}, {Di Mille},
  {Foschini}, {La Mura} et~al.}]{2016cracco1}
{Cracco}, V., {Ciroi}, S., {Berton}, M., {Di Mille}, F., {Foschini}, L., {La
  Mura}, G., et~al. (2016).
\newblock {A spectroscopic analysis of a sample of narrow-line Seyfert 1
  galaxies selected from the Sloan Digital Sky Survey}.
\newblock \emph{\mnras} 462, 1256--1280.
\newblock \doi{10.1093/mnras/stw1689}
\bibAnnoteFile{2016cracco1}

\bibitem[{{Crenshaw} et~al.(2003){Crenshaw}, {Kraemer}, and
  {Gabel}}]{2003crenshaw1}
{Crenshaw}, D.~M., {Kraemer}, S.~B., and {Gabel}, J.~R. (2003).
\newblock {The Host Galaxies of Narrow-Line Seyfert 1 Galaxies: Evidence for
  Bar-Driven Fueling}.
\newblock \emph{\aj} 126, 1690--1698.
\newblock \doi{10.1086/377625}
\bibAnnoteFile{2003crenshaw1}

\bibitem[{{Czerny} et~al.(2009){Czerny}, {Siemiginowska}, {Janiuk},
  {Nikiel-Wroczy{\'n}ski}, and {Stawarz}}]{2009czerny1}
{Czerny}, B., {Siemiginowska}, A., {Janiuk}, A., {Nikiel-Wroczy{\'n}ski}, B.,
  and {Stawarz}, {\L}. (2009).
\newblock {Accretion Disk Model of Short-Timescale Intermittent Activity in
  Young Radio Sources}.
\newblock \emph{\apj} 698, 840--851.
\newblock \doi{10.1088/0004-637X/698/1/840}
\bibAnnoteFile{2009czerny1}

\bibitem[{{D'Ammando} et~al.(2018){D'Ammando}, {Acosta-Pulido}, {Capetti},
  {Baldi}, {Orienti}, {Raiteri} et~al.}]{2018dammando1}
{D'Ammando}, F., {Acosta-Pulido}, J.~A., {Capetti}, A., {Baldi}, R.~D.,
  {Orienti}, M., {Raiteri}, C.~M., et~al. (2018).
\newblock {The host galaxy of the {$\gamma$}-ray-emitting narrow-line Seyfert 1
  galaxy PKS 1502+036}.
\newblock \emph{\mnras} 478, L66--L71.
\newblock \doi{10.1093/mnrasl/sly072}
\bibAnnoteFile{2018dammando1}

\bibitem[{{D'Ammando} et~al.(2017){D'Ammando}, {Acosta-Pulido}, {Capetti},
  {Raiteri}, {Baldi}, {Orienti} et~al.}]{2017dammando1}
{D'Ammando}, F., {Acosta-Pulido}, J.~A., {Capetti}, A., {Raiteri}, C.~M.,
  {Baldi}, R.~D., {Orienti}, M., et~al. (2017).
\newblock {Uncovering the host galaxy of the {$\gamma$}-ray-emitting
  narrow-line Seyfert 1 galaxy FBQS J1644+2619}.
\newblock \emph{\mnras} 469, L11--L15.
\newblock \doi{10.1093/mnrasl/slx042}
\bibAnnoteFile{2017dammando1}

\bibitem[{{Decarli} et~al.(2008){Decarli}, {Dotti}, {Fontana}, and
  {Haardt}}]{2008decarli1}
{Decarli}, R., {Dotti}, M., {Fontana}, M., and {Haardt}, F. (2008).
\newblock {Are the black hole masses in narrow-line Seyfert 1 galaxies actually
  small?}
\newblock \emph{\mnras} 386, L15--L19.
\newblock \doi{10.1111/j.1745-3933.2008.00451.x}
\bibAnnoteFile{2008decarli1}

\bibitem[{{Dennett-Thorpe} et~al.(2002){Dennett-Thorpe}, {Scheuer}, {Laing},
  {Bridle}, {Pooley}, and {Reich}}]{2002dennettthorpe1}
{Dennett-Thorpe}, J., {Scheuer}, P.~A.~G., {Laing}, R.~A., {Bridle}, A.~H.,
  {Pooley}, G.~G., and {Reich}, W. (2002).
\newblock {Jet reorientation in active galactic nuclei: two winged radio
  galaxies}.
\newblock \emph{\mnras} 330, 609--620.
\newblock \doi{10.1046/j.1365-8711.2002.05106.x}
\bibAnnoteFile{2002dennettthorpe1}

\bibitem[{{Deo} et~al.(2006){Deo}, {Crenshaw}, and {Kraemer}}]{2006deo1}
{Deo}, R.~P., {Crenshaw}, D.~M., and {Kraemer}, S.~B. (2006).
\newblock {The Host Galaxies of Narrow-Line Seyfert 1 Galaxies: Nuclear Dust
  Morphology and Starburst Rings}.
\newblock \emph{\aj} 132, 321--346.
\newblock \doi{10.1086/504894}
\bibAnnoteFile{2006deo1}

\bibitem[{{Du} et~al.(2018){Du}, {Brotherton}, {Wang}, {Huang}, {Hu}, {Kasper}
  et~al.}]{2018du1}
{Du}, P., {Brotherton}, M.~S., {Wang}, K., {Huang}, Z.-P., {Hu}, C., {Kasper},
  D.~H., et~al. (2018).
\newblock {Monitoring AGNs with H{\ensuremath{\beta}} Asymmetry. I. First
  Results: Velocity-resolved Reverberation Mapping}.
\newblock \emph{\apj} 869, 142.
\newblock \doi{10.3847/1538-4357/aaed2c}
\bibAnnoteFile{2018du1}

\bibitem[{{Du} et~al.(2014){Du}, {Hu}, {Lu}, {Wang}, {Qiu}, {Li}
  et~al.}]{2014du1}
{Du}, P., {Hu}, C., {Lu}, K.-X., {Wang}, F., {Qiu}, J., {Li}, Y.-R., et~al.
  (2014).
\newblock {Supermassive Black Holes with High Accretion Rates in Active
  Galactic Nuclei. I. First Results from a New Reverberation Mapping Campaign}.
\newblock \emph{\apj} 782, 45.
\newblock \doi{10.1088/0004-637X/782/1/45}
\bibAnnoteFile{2014du1}

\bibitem[{{Ekers} et~al.(1978){Ekers}, {Fanti}, {Lari}, and
  {Parma}}]{1978ekers1}
{Ekers}, R.~D., {Fanti}, R., {Lari}, C., and {Parma}, P. (1978).
\newblock {NGC326 - A radio galaxy with a precessing beam}.
\newblock \emph{\nat} 276, 588--590.
\newblock \doi{10.1038/276588a0}
\bibAnnoteFile{1978ekers1}

\bibitem[{{Foschini}(2017)}]{2017foschini1}
{Foschini}, L. (2017).
\newblock {What we talk about when we talk about blazars?}
\newblock \emph{Frontiers in Astronomy and Space Sciences} 4, 6.
\newblock \doi{10.3389/fspas.2017.00006}
\bibAnnoteFile{2017foschini1}

\bibitem[{{Foschini}(2020)}]{2020foschini1}
{Foschini}, L. (2020).
\newblock {Jetted Narrow-Line Seyfert 1 Galaxies \& Co.: Where Do We Stand?}
\newblock \emph{Universe} 6, 136.
\newblock \doi{10.3390/universe6090136}
\bibAnnoteFile{2020foschini1}

\bibitem[{{Foschini} et~al.(2015){Foschini}, {Berton}, {Caccianiga}, {Ciroi},
  {Cracco}, {Peterson} et~al.}]{2015foschini1}
{Foschini}, L., {Berton}, M., {Caccianiga}, A., {Ciroi}, S., {Cracco}, V.,
  {Peterson}, B.~M., et~al. (2015).
\newblock {Properties of flat-spectrum radio-loud narrow-line Seyfert 1
  galaxies}.
\newblock \emph{\aap} 575, A13.
\newblock \doi{10.1051/0004-6361/201424972}
\bibAnnoteFile{2015foschini1}

\bibitem[{{Fraix-Burnet} et~al.(2017){Fraix-Burnet}, {Marziani}, {D'Onofrio},
  and {Dultzin}}]{2017fraixburnet1}
{Fraix-Burnet}, D., {Marziani}, P., {D'Onofrio}, M., and {Dultzin}, D. (2017).
\newblock {The phylogeny of quasars and the ontogeny of their central black
  holes}.
\newblock \emph{Frontiers in Astronomy and Space Sciences} 4, 1.
\newblock \doi{10.3389/fspas.2017.00001}
\bibAnnoteFile{2017fraixburnet1}

\bibitem[{{Gallo} et~al.(2006){Gallo}, {Edwards}, {Ferrero}, {Kataoka},
  {Lewis}, {Ellingsen} et~al.}]{2006gallo1}
{Gallo}, L.~C., {Edwards}, P.~G., {Ferrero}, E., {Kataoka}, J., {Lewis}, D.~R.,
  {Ellingsen}, S.~P., et~al. (2006).
\newblock {The spectral energy distribution of PKS 2004-447: a compact
  steep-spectrum source and possible radio-loud narrow-line Seyfert 1 galaxy}.
\newblock \emph{\mnras} 370, 245--254.
\newblock \doi{10.1111/j.1365-2966.2006.10482.x}
\bibAnnoteFile{2006gallo1}

\bibitem[{{Goodrich}(1989)}]{1989goodrich1}
{Goodrich}, R.~W. (1989).
\newblock {Spectropolarimetry of 'narrow-line' Seyfert 1 galaxies}.
\newblock \emph{\apj} 342, 224--234.
\newblock \doi{10.1086/167586}
\bibAnnoteFile{1989goodrich1}

\bibitem[{{Gopal-Krishna} et~al.(2012){Gopal-Krishna}, {Biermann}, {Gergely},
  and {Wiita}}]{2012gopalkrishna1}
{Gopal-Krishna}, {Biermann}, P.~L., {Gergely}, L.~{\'A}., and {Wiita}, P.~J.
  (2012).
\newblock {On the origin of X-shaped radio galaxies}.
\newblock \emph{Research in Astronomy and Astrophysics} 12, 127--146.
\newblock \doi{10.1088/1674-4527/12/2/002}
\bibAnnoteFile{2012gopalkrishna1}

\bibitem[{{Hamilton} et~al.(2021){Hamilton}, {Berton}, {Ant{\'o}n}, {Busoni},
  {Caccianiga}, {Ciroi} et~al.}]{2021hamilton1}
{Hamilton}, T.~S., {Berton}, M., {Ant{\'o}n}, S., {Busoni}, L., {Caccianiga},
  A., {Ciroi}, S., et~al. (2021).
\newblock {Observations of the {\ensuremath{\gamma}}-ray-emitting narrow-line
  Seyfert 1, SBS 0846+513, and its host galaxy}.
\newblock \emph{\mnras} 504, 5188--5198.
\newblock \doi{10.1093/mnras/stab1046}
\bibAnnoteFile{2021hamilton1}

\bibitem[{{Hancock} et~al.(2010){Hancock}, {Sadler}, {Mahony}, and
  {Ricci}}]{2010hancock1}
{Hancock}, P.~J., {Sadler}, E.~M., {Mahony}, E.~K., and {Ricci}, R. (2010).
\newblock {Observations and properties of candidate high-frequency GPS radio
  sources in the AT20G survey}.
\newblock \emph{\mnras} 408, 1187--1206.
\newblock \doi{10.1111/j.1365-2966.2010.17199.x}
\bibAnnoteFile{2010hancock1}

\bibitem[{{Hern{\'a}ndez-Garc{\'\i}a} et~al.(2017){Hern{\'a}ndez-Garc{\'\i}a},
  {Panessa}, {Giroletti}, {Ghisellini}, {Bassani}, {Masetti}
  et~al.}]{2017hernandezgarcia1}
{Hern{\'a}ndez-Garc{\'\i}a}, L., {Panessa}, F., {Giroletti}, M., {Ghisellini},
  G., {Bassani}, L., {Masetti}, N., et~al. (2017).
\newblock {Restarting activity in the nucleus of PBC J2333.9-2343. An extreme
  case of jet realignment}.
\newblock \emph{\aap} 603, A131.
\newblock \doi{10.1051/0004-6361/201730530}
\bibAnnoteFile{2017hernandezgarcia1}

\bibitem[{{J{\"a}rvel{\"a}} et~al.(2020){J{\"a}rvel{\"a}}, {Berton}, {Ciroi},
  {Congiu}, {L{\"a}hteenm{\"a}ki}, and {Di Mille}}]{2020jarvela1}
{J{\"a}rvel{\"a}}, E., {Berton}, M., {Ciroi}, S., {Congiu}, E.,
  {L{\"a}hteenm{\"a}ki}, A., and {Di Mille}, F. (2020).
\newblock {SDSS J211852.96-073227.5: The first non-local, interacting,
  late-type intermediate Seyfert galaxy with relativistic jets}.
\newblock \emph{\aap} 636, L12.
\newblock \doi{10.1051/0004-6361/202037826}
\bibAnnoteFile{2020jarvela1}

\bibitem[{{J{\"a}rvel{\"a}} et~al.(2018){J{\"a}rvel{\"a}},
  {L{\"a}hteenm{\"a}ki}, and {Berton}}]{2018jarvela1}
{J{\"a}rvel{\"a}}, E., {L{\"a}hteenm{\"a}ki}, A., and {Berton}, M. (2018).
\newblock {Near-infrared morphologies of the host galaxies of narrow-line
  Seyfert 1 galaxies}.
\newblock \emph{\aap} 619, A69.
\newblock \doi{10.1051/0004-6361/201832876}
\bibAnnoteFile{2018jarvela1}

\bibitem[{{J{\"a}rvel{\"a}} et~al.(2017){J{\"a}rvel{\"a}},
  {L{\"a}hteenm{\"a}ki}, {Lietzen}, {Poudel}, {Hein{\"a}m{\"a}ki}, and
  {Einasto}}]{2017jarvela1}
{J{\"a}rvel{\"a}}, E., {L{\"a}hteenm{\"a}ki}, A., {Lietzen}, H., {Poudel}, A.,
  {Hein{\"a}m{\"a}ki}, P., and {Einasto}, M. (2017).
\newblock {Large-scale environments of narrow-line Seyfert 1 galaxies}.
\newblock \emph{\aap} 606, A9.
\newblock \doi{10.1051/0004-6361/201731318}
\bibAnnoteFile{2017jarvela1}

\bibitem[{{Komatsu} et~al.(2011){Komatsu}, {Smith}, {Dunkley}, {Bennett},
  {Gold}, {Hinshaw} et~al.}]{2011komatsu1}
{Komatsu}, E., {Smith}, K.~M., {Dunkley}, J., {Bennett}, C.~L., {Gold}, B.,
  {Hinshaw}, G., et~al. (2011).
\newblock {Seven-year Wilkinson Microwave Anisotropy Probe (WMAP) Observations:
  Cosmological Interpretation}.
\newblock \emph{\apjs} 192, 18.
\newblock \doi{10.1088/0067-0049/192/2/18}
\bibAnnoteFile{2011komatsu1}

\bibitem[{{Komossa} et~al.(2006){Komossa}, {Voges}, {Xu}, {Mathur}, {Adorf},
  {Lemson} et~al.}]{2006komossa1}
{Komossa}, S., {Voges}, W., {Xu}, D., {Mathur}, S., {Adorf}, H.-M., {Lemson},
  G., et~al. (2006).
\newblock {Radio-loud Narrow-Line Type 1 Quasars}.
\newblock \emph{\aj} 132, 531--545.
\newblock \doi{10.1086/505043}
\bibAnnoteFile{2006komossa1}

\bibitem[{{Kotilainen} et~al.(2016){Kotilainen}, {Le{\'o}n-Tavares},
  {Olgu{\'{\i}}n-Iglesias}, {Baes}, {An{\'o}rve}, {Chavushyan}
  et~al.}]{2016kotilainen1}
{Kotilainen}, J.~K., {Le{\'o}n-Tavares}, J., {Olgu{\'{\i}}n-Iglesias}, A.,
  {Baes}, M., {An{\'o}rve}, C., {Chavushyan}, V., et~al. (2016).
\newblock {Discovery of a Pseudobulge Galaxy Launching Powerful Relativistic
  Jets}.
\newblock \emph{\apj} 832, 157.
\newblock \doi{10.3847/0004-637X/832/2/157}
\bibAnnoteFile{2016kotilainen1}

\bibitem[{{Krongold} et~al.(2001){Krongold}, {Dultzin-Hacyan}, and
  {Marziani}}]{2001krongold1}
{Krongold}, Y., {Dultzin-Hacyan}, D., and {Marziani}, P. (2001).
\newblock {Host Galaxies and Circumgalactic Environment of ``Narrow Line''
  Seyfert 1 Nuclei}.
\newblock \emph{\aj} 121, 702--709.
\newblock \doi{10.1086/318768}
\bibAnnoteFile{2001krongold1}

\bibitem[{{L{\"a}hteenm{\"a}ki} et~al.(2017){L{\"a}hteenm{\"a}ki},
  {J{\"a}rvel{\"a}}, {Hovatta}, {Tornikoski}, {Harrison}, {L{\'o}pez-Caniego}
  et~al.}]{2017lahteenmaki1}
{L{\"a}hteenm{\"a}ki}, A., {J{\"a}rvel{\"a}}, E., {Hovatta}, T., {Tornikoski},
  M., {Harrison}, D.~L., {L{\'o}pez-Caniego}, M., et~al. (2017).
\newblock {37 GHz observations of narrow-line Seyfert 1 galaxies}.
\newblock \emph{\aap} 603, A100.
\newblock \doi{10.1051/0004-6361/201630257}
\bibAnnoteFile{2017lahteenmaki1}

\bibitem[{{L{\"a}hteenm{\"a}ki} et~al.(2018){L{\"a}hteenm{\"a}ki},
  {J{\"a}rvel{\"a}}, {Ramakrishnan}, {Tornikoski}, {Tammi}, {Vera}
  et~al.}]{2018lahteenmaki1}
{L{\"a}hteenm{\"a}ki}, A., {J{\"a}rvel{\"a}}, E., {Ramakrishnan}, V.,
  {Tornikoski}, M., {Tammi}, J., {Vera}, R.~J.~C., et~al. (2018).
\newblock {Radio jets and gamma-ray emission in radio-silent narrow-line
  Seyfert 1 galaxies}.
\newblock \emph{\aap} 614, L1.
\newblock \doi{10.1051/0004-6361/201833378}
\bibAnnoteFile{2018lahteenmaki1}

\bibitem[{{Laor}(2000)}]{2000laor1}
{Laor}, A. (2000).
\newblock {On Black Hole Masses and Radio Loudness in Active Galactic Nuclei}.
\newblock \emph{\apjl} 543, L111--L114.
\newblock \doi{10.1086/317280}
\bibAnnoteFile{2000laor1}

\bibitem[{{Longinotti} et~al.(2018){Longinotti}, {Vega}, {Krongold},
  {Aretxaga}, {Yun}, {Chavushyan} et~al.}]{2018longinotti1}
{Longinotti}, A.~L., {Vega}, O., {Krongold}, Y., {Aretxaga}, I., {Yun}, M.,
  {Chavushyan}, V., et~al. (2018).
\newblock {Early Science with the Large Millimeter Telescope: An Energy-driven
  Wind Revealed by Massive Molecular and Fast X-Ray Outflows in the Seyfert
  Galaxy IRAS 17020+4544}.
\newblock \emph{\apjl} 867, L11.
\newblock \doi{10.3847/2041-8213/aae5fd}
\bibAnnoteFile{2018longinotti1}

\bibitem[{{Mathur}(2000)}]{2000mathur1}
{Mathur}, S. (2000).
\newblock {Narrow-line Seyfert 1 galaxies and the evolution of galaxies and
  active galaxies}.
\newblock \emph{\mnras} 314, L17--L20.
\newblock \doi{10.1046/j.1365-8711.2000.03530.x}
\bibAnnoteFile{2000mathur1}

\bibitem[{{O'Dea} and {Saikia}(2021)}]{2021odea1}
{O'Dea}, C.~P. and {Saikia}, D.~J. (2021).
\newblock {Compact steep-spectrum and peaked-spectrum radio sources}.
\newblock \emph{\aapr} 29, 3.
\newblock \doi{10.1007/s00159-021-00131-w}
\bibAnnoteFile{2021odea1}

\bibitem[{{Ohta} et~al.(2007){Ohta}, {Aoki}, {Kawaguchi}, and
  {Kiuchi}}]{2007ohta1}
{Ohta}, K., {Aoki}, K., {Kawaguchi}, T., and {Kiuchi}, G. (2007).
\newblock {A Bar Fuels a Supermassive Black Hole?: Host Galaxies of Narrow-Line
  Seyfert 1 Galaxies}.
\newblock \emph{\apjs} 169, 1--20.
\newblock \doi{10.1086/510204}
\bibAnnoteFile{2007ohta1}

\bibitem[{{Olgu{\'\i}n-Iglesias} et~al.(2020){Olgu{\'\i}n-Iglesias},
  {Kotilainen}, and {Chavushyan}}]{2020olguiniglesias1}
{Olgu{\'\i}n-Iglesias}, A., {Kotilainen}, J., and {Chavushyan}, V. (2020).
\newblock {The disc-like host galaxies of radio-loud narrow-line Seyfert 1s}.
\newblock \emph{\mnras} 492, 1450--1464.
\newblock \doi{10.1093/mnras/stz3549}
\bibAnnoteFile{2020olguiniglesias1}

\bibitem[{{Orban de Xivry} et~al.(2011){Orban de Xivry}, {Davies},
  {Schartmann}, {Komossa}, {Marconi}, {Hicks} et~al.}]{2011orbandexivry1}
{Orban de Xivry}, G., {Davies}, R., {Schartmann}, M., {Komossa}, S., {Marconi},
  A., {Hicks}, E., et~al. (2011).
\newblock {The role of secular evolution in the black hole growth of
  narrow-line Seyfert 1 galaxies}.
\newblock \emph{\mnras} 417, 2721--2736.
\newblock \doi{10.1111/j.1365-2966.2011.19439.x}
\bibAnnoteFile{2011orbandexivry1}

\bibitem[{{Oshlack} et~al.(2001){Oshlack}, {Webster}, and
  {Whiting}}]{2001oshlack1}
{Oshlack}, A.~Y.~K.~N., {Webster}, R.~L., and {Whiting}, M.~T. (2001).
\newblock {A Very Radio Loud Narrow-Line Seyfert 1: PKS 2004-447}.
\newblock \emph{\apj} 558, 578--582.
\newblock \doi{10.1086/322299}
\bibAnnoteFile{2001oshlack1}

\bibitem[{{Osterbrock} and {Pogge}(1985)}]{1985osterbrock1}
{Osterbrock}, D.~E. and {Pogge}, R.~W. (1985).
\newblock {The spectra of narrow-line Seyfert 1 galaxies}.
\newblock \emph{\apj} 297, 166--176.
\newblock \doi{10.1086/163513}
\bibAnnoteFile{1985osterbrock1}

\bibitem[{{Parma} et~al.(1985){Parma}, {Ekers}, and {Fanti}}]{1985parma1}
{Parma}, P., {Ekers}, R.~D., and {Fanti}, R. (1985).
\newblock {High resolution radio observations of low luminosity radio
  galaxies.}
\newblock \emph{\aaps} 59, 511--521
\bibAnnoteFile{1985parma1}

\bibitem[{{Peterson}(2011)}]{2011peterson1}
{Peterson}, B.~M. (2011).
\newblock {Masses of Black Holes in Active Galactic Nuclei: Implications for
  NLS1s}.
\newblock \emph{ArXiv e-prints}
\bibAnnoteFile{2011peterson1}

\bibitem[{{Rakshit} et~al.(2021){Rakshit}, {Schramm}, {Stalin}, {Tanaka},
  {Paliya}, {Pal} et~al.}]{2021rakshit1}
{Rakshit}, S., {Schramm}, M., {Stalin}, C.~S., {Tanaka}, I., {Paliya}, V.~S.,
  {Pal}, I., et~al. (2021).
\newblock {TXS 1206+549: a new {\ensuremath{\gamma}}-ray detected narrow-line
  Seyfert 1 galaxy at redshift 1.34?}
\newblock \emph{\mnras} \doi{10.1093/mnrasl/slab031}
\bibAnnoteFile{2021rakshit1}

\bibitem[{{Romano} et~al.(2018){Romano}, {Vercellone}, {Foschini}, {Tavecchio},
  {Landoni}, and {Kn{\"o}dlseder}}]{2018romano1}
{Romano}, P., {Vercellone}, S., {Foschini}, L., {Tavecchio}, F., {Landoni}, M.,
  and {Kn{\"o}dlseder}, J. (2018).
\newblock {Prospects for gamma-ray observations of narrow-line Seyfert 1
  galaxies with the Cherenkov Telescope Array}.
\newblock \emph{\mnras} 481, 5046--5061.
\newblock \doi{10.1093/mnras/sty2484}
\bibAnnoteFile{2018romano1}

\bibitem[{{Sani} et~al.(2010){Sani}, {Lutz}, {Risaliti}, {Netzer}, {Gallo},
  {Trakhtenbrot} et~al.}]{2010sani1}
{Sani}, E., {Lutz}, D., {Risaliti}, G., {Netzer}, H., {Gallo}, L.~C.,
  {Trakhtenbrot}, B., et~al. (2010).
\newblock {Enhanced star formation in narrow-line Seyfert 1 active galactic
  nuclei revealed by Spitzer}.
\newblock \emph{\mnras} 403, 1246--1260.
\newblock \doi{10.1111/j.1365-2966.2009.16217.x}
\bibAnnoteFile{2010sani1}

\bibitem[{{Sulentic} et~al.(2000){Sulentic}, {Zwitter}, {Marziani}, and
  {Dultzin-Hacyan}}]{2000sulentic1}
{Sulentic}, J.~W., {Zwitter}, T., {Marziani}, P., and {Dultzin-Hacyan}, D.
  (2000).
\newblock {Eigenvector 1: An Optimal Correlation Space for Active Galactic
  Nuclei}.
\newblock \emph{\apjl} 536, L5--L9.
\newblock \doi{10.1086/312717}
\bibAnnoteFile{2000sulentic1}

\bibitem[{{van Haarlem} et~al.(2013){van Haarlem}, {Wise}, {Gunst}, {Heald},
  {McKean}, {Hessels} et~al.}]{2013vanhaarlem1}
{van Haarlem}, M.~P., {Wise}, M.~W., {Gunst}, A.~W., {Heald}, G., {McKean},
  J.~P., {Hessels}, J.~W.~T., et~al. (2013).
\newblock {LOFAR: The LOw-Frequency ARray}.
\newblock \emph{\aap} 556, A2.
\newblock \doi{10.1051/0004-6361/201220873}
\bibAnnoteFile{2013vanhaarlem1}

\bibitem[{{Wang} et~al.(2016){Wang}, {Du}, {Hu}, {Bai}, {Wang}, {Yi}
  et~al.}]{2016wang1}
{Wang}, F., {Du}, P., {Hu}, C., {Bai}, J.-M., {Wang}, C.-J., {Yi}, W.-M.,
  et~al. (2016).
\newblock {Reverberation Mapping of the Gamma-Ray Loud Narrow-line Seyfert 1
  Galaxy 1H 0323+342}.
\newblock \emph{\apj} 824, 149.
\newblock \doi{10.3847/0004-637X/824/2/149}
\bibAnnoteFile{2016wang1}

\bibitem[{{Yao} and {Komossa}(2021)}]{2021yao1}
{Yao}, S. and {Komossa}, S. (2021).
\newblock {Spectroscopic classification, variability, and SED of the
  Fermi-detected CSS 3C 286: the radio-loudest NLS1 galaxy?}
\newblock \emph{\mnras} 501, 1384--1393.
\newblock \doi{10.1093/mnras/staa3708}
\bibAnnoteFile{2021yao1}

\bibitem[{{Yuan} et~al.(2008){Yuan}, {Zhou}, {Komossa}, {Dong}, {Wang}, {Lu}
  et~al.}]{2008yuan1}
{Yuan}, W., {Zhou}, H.~Y., {Komossa}, S., {Dong}, X.~B., {Wang}, T.~G., {Lu},
  H.~L., et~al. (2008).
\newblock {A Population of Radio-Loud Narrow-Line Seyfert 1 Galaxies with
  Blazar-Like Properties?}
\newblock \emph{\apj} 685, 801--827.
\newblock \doi{10.1086/591046}
\bibAnnoteFile{2008yuan1}

\end{thebibliography}


\section*{Figure captions}

\begin{table}[h!]
\caption[]{Summary of the source properties, and archival average Metsähovi flux densities. Columns: (1) source name, (2) and (3) right ascension and declination, from the JVLA X or C band observations in case of a detection, otherwise from FIRST (marked after the source name with $^{X}$, $^{C}$, or $^{F}$, respectively), (4) redshift, obtained from the Sloan Digital Sky Survey \citep{2009abazajian1}, (5) black hole mass, (6) scale at the redshift of the source, (7) Metsähovi 37~GHz average flux density.}
\centering
\begin{tabular}{l l l l l l l}
\hline\hline
source                    & RA           & Dec          & $z$   & log $M_{\text{BH}}$ & scale  & $S_{\text{MH, 37~GHz}}^{\text{ave}}$ \\
                          & (hh mm ss.s) & (dd mm ss.s) &       & ($M_\odot$) & (kpc/arcsec) & (mJy) \\ \hline
J102906.69+555625.2$^{F}$ & 10 29 06.69  & +55 56 25.25 & 0.451 & 7.33 & 9.685 & 420 \\ 
J122844.81+501751.2$^{X}$ & 12 28 44.82  & +50 17 51.24 & 0.262 & 6.84 & 5.657 & 410 \\ 
J123220.11+495721.8$^{C}$ & 12 32 20.11  & +49 57 21.79 & 0.262 & 7.30 & 5.626 & 460 \\ 
J150916.18+613716.7$^{F}$ & 15 09 16.17  & +61 37 16.80 & 0.201 & 6.66 & 4.313 & 670 \\ 
J151020.06+554722.0$^{X}$ & 15 10 20.05  & +55 47 22.11 & 0.150 & 6.67 & 3.214 & 450 \\ 
J152205.41+393441.3$^{X}$ & 15 22 05.50  & +39 34 40.46 & 0.077 & 5.97 & 1.650 & 590 \\ 
J164100.10+345452.7$^{X}$ & 16 41 00.10  & +34 54 52.66 & 0.164 & 7.15 & 3.518 & 370 \\ \hline
\end{tabular} 
\label{tab:sample}
\end{table}

\begin{table}[h!]
\caption[]{Summary of the radio flux densities of the sources. Columns: (1) source name, (2) units, (3) JVLA flux densities and rms at 1.6~GHz, (4) JVLA flux densities and rms at 5.2~GHz, (5) JVLA flux densities and rms at 9.0~GHz, (6) LoTTS flux densities at 144~MHz. N/A means the source in not within the published LoTTS region.}
\centering
\begin{tabular}{l l l l l l}
\hline\hline
source     &                        & $S_{\text{1.6~GHz, JVLA}}$ & $S_{\text{5.2~GHz, JVLA}}$ & $S_{\text{9.0~GHz, JVLA}}$ & $S_{\text{144~MHz, LoTTS}}$  \\ \hline
J1228+5017 & rms (mJy)              & 0.020                      & 0.008                      & 0.007 &                                                   \\
           & int (mJy)              & 0.97 $\pm$ 0.04            & 0.29 $\pm$ 0.01            & 0.19 $\pm$ 0.01            & 2.3 $\pm$ 0.1                \\ 
           & peak (mJy beam$^{-1}$) & 0.73 $\pm$ 0.02            & 0.24 $\pm$ 0.01            & 0.18 $\pm$ 0.01            & 2.1 $\pm$ 0.1                \\
           &                        &                            &                            &                            &                              \\
J1522+3934 & rms (mJy)              & 0.017                      & 0.010                      & 0.007                      &                              \\
           & int (mJy)              & 3.18 $\pm$ 0.07            & 0.40 $\pm$ 0.02            & 0.35 $\pm$ 0.02            & N/A                          \\
           & peak (mJy beam$^{-1}$) & 1.05 $\pm$ 0.05            & 0.33 $\pm$ 0.01            & 0.19 $\pm$ 0.01            &                              \\
           &                        &                            &                            &                            &                              \\
J1641+3454 & rms (mJy)              & 0.025                      & 0.007                      & 0.007                      &                              \\
           & int (mJy)              & 2.34 $\pm$ 0.06            & 0.69 $\pm$ 0.02            & 0.35 $\pm$ 0.01            & N/A                          \\ 
           & peak (mJy beam$^{-1}$) & 1.82 $\pm$ 0.03            & 0.51 $\pm$ 0.01            & 0.28 $\pm$ 0.01            &                              \\ 

\hline
\end{tabular} 
\label{tab:flux}
\end{table}

\begin{table}[ht!]
\caption[]{Weighted total and core spectral indices measured from the $\alpha$ maps. Columns: (1) Source name, (2) are within which the spectral index was measured, (3) 1.6~GHz spectral index, (4) 5.2~GHz spectral index; (5) 9.0~GHz spectral index.} \label{tab:spinds}
\centering
\begin{tabular}{l l l l l}
\hline\hline
source       & area  & $\alpha_{\text{1.6~GHz, JVLA}}$ & $\alpha_{\text{5.2~GHz, JVLA}}$ & $\alpha_{\text{9.0~GHz, JVLA}}$ \\
             &       &                        &                        &  \\ \hline
J1228+5017   & total & -0.41 $\pm$ 0.01       & -1.00 $\pm$	0.02       & -0.31 $\pm$ 0.01 \\
             & core  & -0.38 $\pm$ 0.01       & -1.24 $\pm$ 0.06       & -0.43 $\pm$ 0.03 \\
J1522+3934   & total & -0.80 $\pm$ 0.01       & -0.42 $\pm$ 0.01       &  0.49 $\pm$ 0.01 \\
             & core  & -0.71 $\pm$ 0.01       & -0.47 $\pm$ 0.02       &  0.35 $\pm$ 0.01 \\
J1641+3454   & total & -0.85 $\pm$ 0.01       & -0.61 $\pm$ 0.01       & -0.83 $\pm$ 0.01 \\
             & core  & -0.99 $\pm$ 0.01       & -0.78 $\pm$ 0.01       & -1.16 $\pm$ 0.01 \\ \hline
\end{tabular}  
\label{tab:j2118}
\end{table}

\begin{table}[ht!]
\caption[]{Traditional spectral indices calculated using the peak flux densities. Columns: (1) Source name, (2) In-band spectral index with a central frequency of 1.6~GHz, band-width 1~GHz, (3) in-band spectral index with a central frequency of 5.2~GHz, band-width 2~GHz, (4) in-band spectral index with a central frequency of 9.0~GHz, band-width 2~GHz.} \label{tab:tradspind}
\centering
\begin{tabular}{l l l l}
\hline\hline
source       &  $\alpha_{\text{1.6~GHz, JVLA}}$ & $\alpha_{\text{5.2~GHz, JVLA}}$ & $\alpha_{\text{9.0~GHz, JVLA}}$ \\
             &                        &                        &  \\ \hline
J1228+5017   & -0.19 $\pm$ 0.19       & -1.11 $\pm$ 0.43       & -0.46 $\pm$ 0.91 \\
J1522+3934   & -0.48 $\pm$ 0.30       & -0.55 $\pm$ 0.45       & -0.46 $\pm$ 1.08 \\
J1641+3454   & -0.88 $\pm$ 0.13       & -0.94 $\pm$ 0.22       & -1.13 $\pm$ 0.52 \\ \hline
\end{tabular} 
\label{tab:j2118}
\end{table}

\end{document}